\documentclass[aps,prl,twocolumn,showpacs,amsmath,amssymb,superscriptaddress]{revtex4-1}

\usepackage{graphicx}
\usepackage{dcolumn}
\usepackage{bm}
\usepackage{epsfig}
\usepackage{epstopdf}
\usepackage{float}	
\usepackage{color}
\usepackage[title]{appendix}
\usepackage{mathrsfs}
\usepackage{enumitem}
\usepackage[hidelinks]{hyperref}
\usepackage{braket}\usepackage{mathtools}
\usepackage{amsmath}

\usepackage{tabularx}
\usepackage{float}
\newcolumntype{C}{>{\centering\arraybackslash}X}

\begin{document}
\title{Nonreciprocal Control of the Speed of Light Using Cavity Magnonics}

\author{Jiguang~Yao}
\affiliation{Department of Physics and Astronomy, University of Manitoba, Winnipeg, Canada R3T 2N2}

\author{Chenyang~Lu}
\affiliation{Department of Physics and Astronomy, University of Manitoba, Winnipeg, Canada R3T 2N2}

\author{Xiaolong~Fan}
\affiliation{The Key Laboratory for Magnetism and Magnetic Materials of the Ministry of Education, Lanzhou University, Lanzhou 730000, China}

\author{Desheng~Xue}
\affiliation{The Key Laboratory for Magnetism and Magnetic Materials of the Ministry of Education, Lanzhou University, Lanzhou 730000, China}

\author{Greg E. Bridges}
\affiliation{Department of Electrical and Computer Engineering, University of Manitoba, Winnipeg, Canada R3T 5V6}

\author{C.-M.~Hu} \email{hu@physics.umanitoba.ca; \\URL: http://www.physics.umanitoba.ca/$\sim$hu}
\affiliation{Department of Physics and Astronomy, University of Manitoba, Winnipeg, Canada R3T 2N2}

\date{\today}

\begin{abstract}
	
We demonstrate nonreciprocal control of the speed of light by sending a microwave pulse through a cavity magnonics device. In contrast to reciprocal group velocity controlled by conventional electromagnetically induced transparency (EIT) effect, incorporating dissipative magnon-photon coupling establishes a non-reciprocal EIT effect, allowing slow and fast light propagation in opposite directions at the same frequency with comparable amplitude. Remarkably, reversing the magnetic field enables a directional switch between non-reciprocal fast and slow light. This discovery may offer new possibilities for pulse time regulation in microwave signal communications, neuromorphic computing, and quantum signal processing.
\end{abstract}

\maketitle

\textit{Introduction.--}
Non-reciprocal control of light is a critical research area with broad implications for telecommunications \cite{reiskarimian2016magnetic}, photonic circuits \cite{feng2011nonreciprocal}, optical networks \cite{shen2023nonreciprocal}, and quantum computing \cite{rosario2018nonreciprocity}. Past efforts focused on achieving directional control of the amplitude (or power) of electromagnetic waves, driven by the need for conventional non-reciprocal devices such as isolators and circulators \cite{reiskarimian2016magnetic,feng2011nonreciprocal,shen2023nonreciprocal,rosario2018nonreciprocity}. These devices allow one-way wave propagation, preventing reflections and ensuring interference-free signal processing and quantum measurements.

\begin{figure} [!ht]
	\begin{center}
		\includegraphics[width=8.8 cm]{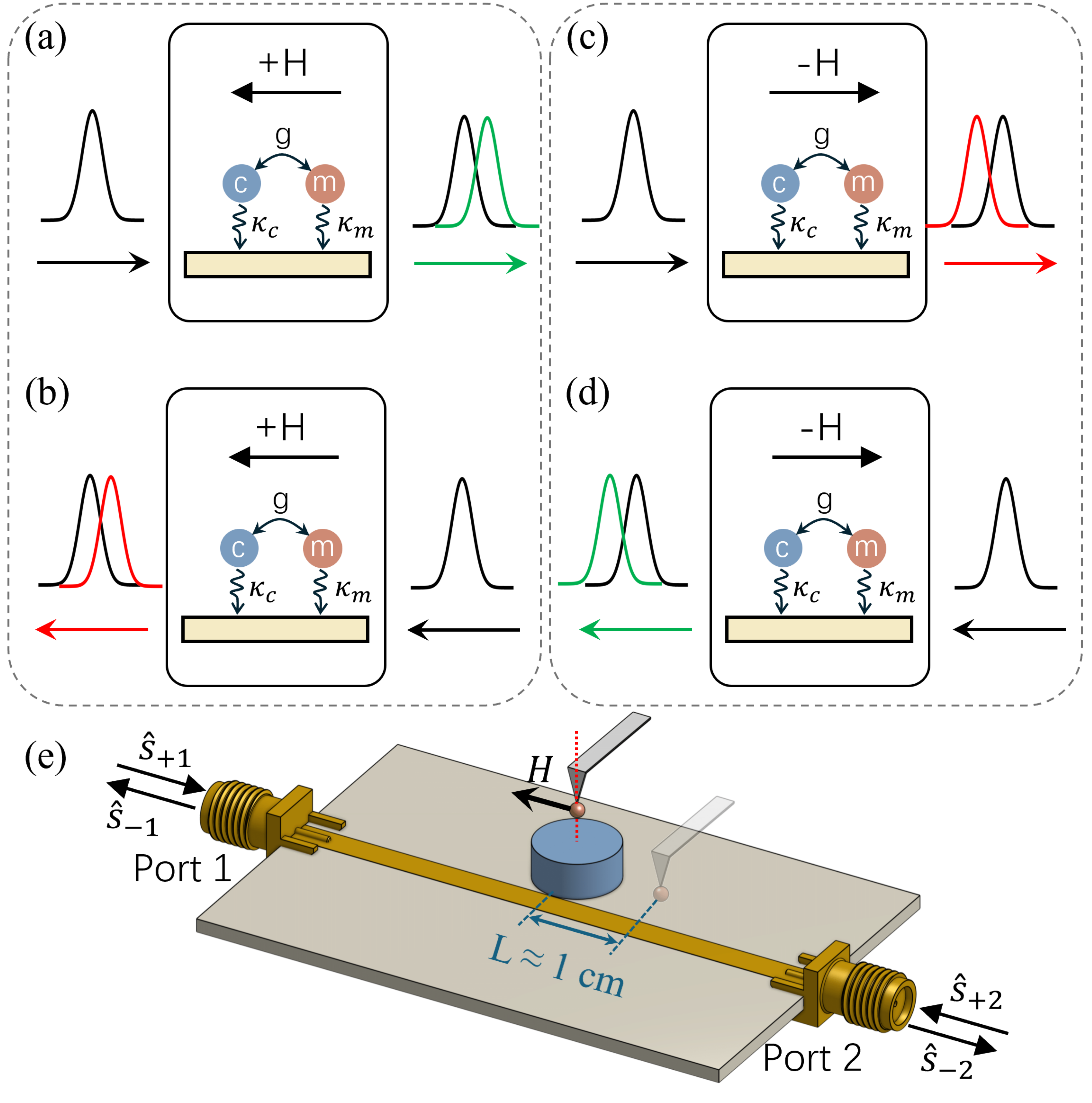}
		\caption{(a)-(d) Schematic of nonreciprocal control of the speed of light using a cavity mode $\hat{c}$ and a magnon mode $\hat{m}$. They are coupled to each other with a direct coupling strength $g$, and coupled to a common microstrip with coupling rates $\kappa_c$ and $\kappa_m$, respectively. The left and right panels compare the cases at positive and negative magnetic fields, respectively. The light propagation when the pulse transfers from (a),(c) port 1 to port 2, compared to that from (b),(d) port 2 to port 1. The black curves denote the original pulse through a path without the coupled system, while the green and red curves represent the advanced and delayed pulse through the coupled system, respectively. (e) Schematic of the experimental setup. The coupled system includes a YIG sphere and a DR placed on a microstrip. The solid and the  translucent YIG spheres represent the configurations for the EIT analog at $\kappa_m=0$ and non-reciprocal EIT at $\kappa_m \ne 0$, respectively. The former is placed above the center of DR (along the red dotted line) far from the microstrip, while the latter is alongside the DR near the microstrip, with the center-to-center distance $L \approx 1$ cm. The system is measured by the input $\hat{s}_{+1}$,$\hat{s}_{+2}$ and the output $\hat{s}_{-1}$,$\hat{s}_{-2}$ fields. }
		\label{Fig1}
	\end{center}
\end{figure}

On a different route, significant efforts have also focused on controlling the phase of light pulses, leading to the discovery of slow \cite{hau1999light,bigelow2003observation,totsuka2007slow,el2010observation,safavi2011electromagnetically,heeg2015tunable} and fast light \cite{wang2000gain,bigelow2003superluminal,goyon2021slow}, characterized by positive and negative group delays, respectively. Counter intuitively, slow light can have a group velocity approaching zero, while fast light can exceed the speed of light, $c$, and even become negative. These effects arise from classical interference between different frequency components of the pulse, which do not violate the principle of causality in signal propagation. They have enabled optical buffers and storage \cite{phillips2001storage,liu2001observation,heinze2013stopped}, advanced optical networks \cite{lake2021processing}, and enhanced the effects of light-matter interactions \cite{franke2011rotary,huet2016millisecond,xu2024slow}.

Thus far, these two approaches have remained largely separate. Unlike directional control of light's amplitude, non-reciprocal control of light's speed is much more challenging. Let's take the electromagnetically induced transparency (EIT) as an example \cite{boller1991observation,fleischhauer2005electromagnetically}, which is the key method for producing slow and fast light. Arising from destructive interference between different pathways in atomic \cite{fleischhauer2005electromagnetically} or coupled resonator systems \cite{smith2004coupled,liu2017electromagnetically}, EIT generates the sharp dispersion required for achieving a large group delay while maintaining significant transmission. Breaking the reciprocity of group velocity is achievable through phase modulation, as demonstrated in cavity optomechanical systems \cite{shen2016experimental} and whispering-gallery microresonators \cite{dong2015brillouin}. However, maintaining significantly non-reciprocal group velocities with comparable bi-directional amplitudes is difficult, since the real and imaginary parts of optical response functions are linked by the Kramers-Kronig relations. It remains unclear whether it is feasible to create a non-reciprocal system where slow and fast light can propagate in opposite directions at the same frequency with comparable amplitudes.

In this letter, we address this question by developing a non-reciprocal EIT technique based on cavity magnonics \cite{rameshti2022cavity}. Built on magnetic materials that can break the time-reversal symmetry, cavity magnonics has been used for studying, separately, EIT analog \cite{zhang2014strongly,liu2019room,zhao2021phase,xiong2022tunable}; and non-reciprocal microwave transmissions \cite{wang2019nonreciprocity, zhang2020broadband,li2023unidirectional}. Furthermore, as a versatile platform involving various coupling mechanisms \cite{harder2018level,boventer2019steering,lachance2019hybrid,wang2020dissipative,li2020hybrid,harder2021coherent, yuan2022quantum}, it offers the flexility for engineering light-matter interactions. Capitalizing on these advantages, we achieve unprecedented non-reciprocal control of microwave pulses. As highlighted in Fig. 1(a)-(d), we demonstrate that by sending a microwave pulse to a cavity magnonics device, the pulse advances in time by a factor of over 500 in one propagating direction, but the same pulse delays by a similar amount of time in the opposite direction. Furthermore, the transmission amplitudes are similar along both directions. Most remarkably, such a non-reciprocal fast/slow light can be switched on demand on a compact solid-state platform by simply reversing the applied magnetic field.

\textit{Setup.--}
Our device, as shown by the schematic in Fig. \ref{Fig1}(e), consists of a 4.65 mm-wide, 6 cm-long microstrip with an effective permittivity of 2.2, a yttrium iron garnet (YIG) sphere with 1-mm diameter, and a dielectric resonator (DR) with a diameter of $9.1$ mm, a height of $3.7$ mm and a dielectric constant of $34$. The YIG sphere is biased by an external magnetic field $H$ applied parallel to the microstrip, which controls the resonant frequency $\omega_m/2\pi = \gamma_e \mu_0 (|H|+H_A)$ of the Kittel magnon mode.  Here, $\mu_0$ is the vacuum permeability and $H_A$ is the anisotropy field. From calibrations \cite{Sup}, we get the gyromagnetic ratio $\gamma_e/2\pi = 28$ GHz/T, and the intrinsic magnon damping rate $\alpha_0/2\pi$ = 0.8 MHz. The magnon mode is directly coupled via the dipole interaction with the TE$_{01}$ mode of the DR. This cavity mode radiates microwaves to the open space at $\omega_c/2\pi=6.2035$ GHz with an intrinsic damping rate of $\beta_0/2\pi=17.0$ MHz. The DR is set in the position where the cavity mode is a standing-wave mode under-coupled with the microstrip \cite{yang2024anomalous}, with a fixed extrinsic damping rate $\kappa_c/2\pi=20.5$ MHz. The position of the YIG sphere, which determines the extrinsic damping rate $\kappa_m$ of the magnon mode as well as the direct magnon-photon coupling rate $g$, is controlled by a 3D displacement cantilever. In Fig. \ref{Fig1}(e), the solid and the translucent YIG positions represent two configurations to obtain EIT-anolog and nonreciprocal EIT, respectively, as we demonstrate below in two transmission experiments.

\textit{EIT analog at $\kappa_m$ \textsc{= 0}.--} We first set the device in EIT-analog configuration by placing the YIG sphere above the center of DR far from the microstrip. In this configuration, $\kappa_m = $ 0, and $g$ is controlled by the distance between the DR and the YIG sphere along the dotted red line shown in Fig. \ref{Fig1}(e). Transmission measurements are performed by connecting the microstrip to a vector network analyzer (VNA). The device is probed by the input fields $\hat{s}_{+1}$, $\hat{s}_{+2}$ and the output fields $\hat{s}_{-1}$, $\hat{s}_{-2}$, at the frequency $\omega$.

Using coupled-mode theory \cite{haus1984waves,fan2003temporal}, the equations of motion of the cavity mode $\hat{c}$ and the magnon mode $\hat{m}$, supplemented with the input-output relation, are given by,
\begin{equation}\label{coupled equation of motion1}
	\begin{split}
		\begin{pmatrix} \dot{\hat{c}} \\ \dot{\hat{m}} \end{pmatrix}&=-i \begin{pmatrix} \tilde{\omega}_c-i\kappa_c & g \\ g & \tilde{\omega}_m \end{pmatrix} \begin{pmatrix} \hat{c} \\ \hat{m} \end{pmatrix} -i \begin{pmatrix} \sqrt{\kappa_c} \\ 0 \end{pmatrix} \hat{s}_{+1}, \\
		\hat{s}_{-2}&=\hat{s}_{+1}-i\sqrt{\kappa_c}\hat{c},\\
	\end{split}
\end{equation}
where $\tilde{\omega}_m=\omega_m-i\alpha_0$ and $\tilde{\omega}_c=\omega_c-i\beta_0$. Using Eq. \ref{coupled equation of motion1}, we derive the transmission coefficients \cite{Sup}:
\begin{equation}\label{EIT S21}
	S_{21}(\omega)=S_{12}(\omega)=\frac{(\omega-\tilde{\omega}_m)(\omega-\tilde{\omega}_c)-g^2}{(\omega-\tilde{\omega}_m)(\omega-\tilde{\omega}_c+i\kappa_c)-g^2}.
\end{equation}
The device is reciprocal in this configuration, so we only display the results from $S_{21}$ in this section, and we define the probing frequency detuning $\Delta_c=\omega-\omega_c$ and the field detuning $\Delta_m=\omega_m(H)-\omega_c$. The group delay $\tau_g(\omega)$ is defined as $\tau_g(\omega)=\partial \angle S_{21}(\omega)/\partial \omega$, where $\angle S_{21}$ is the phase of $S_{21}$. Specially at $\Delta_c = \Delta_m = 0$, $\tau_g^0$ is analytically derived \cite{Sup} from Eq. \ref{EIT S21}:
   \begin{equation}\label{group delay as function of g}
   	\tau_g^0 \approx \frac{\kappa_c(g^2-\alpha_0^2)}{[(\kappa_c+\beta_0)\alpha_0+g^2](\alpha_0\beta_0+g^2)},
   \end{equation}
showing that the group delay is controlled by $\alpha_0$, $\beta_0$, $\kappa_c$, and $g$. Eq. \ref{group delay as function of g} shows that to achieve a large group delay, high-quality YIG with a small damping $\alpha_0$ is essential. Similarly, a high-quality resonator with a small $\beta_0$ is beneficial. In contrast, the preferred rate of the coupling $g$ needs being carefully engineered, as will be shown in the experiment.

\begin{figure} [!t]
	\begin{center}
		\includegraphics[width=8.8 cm]{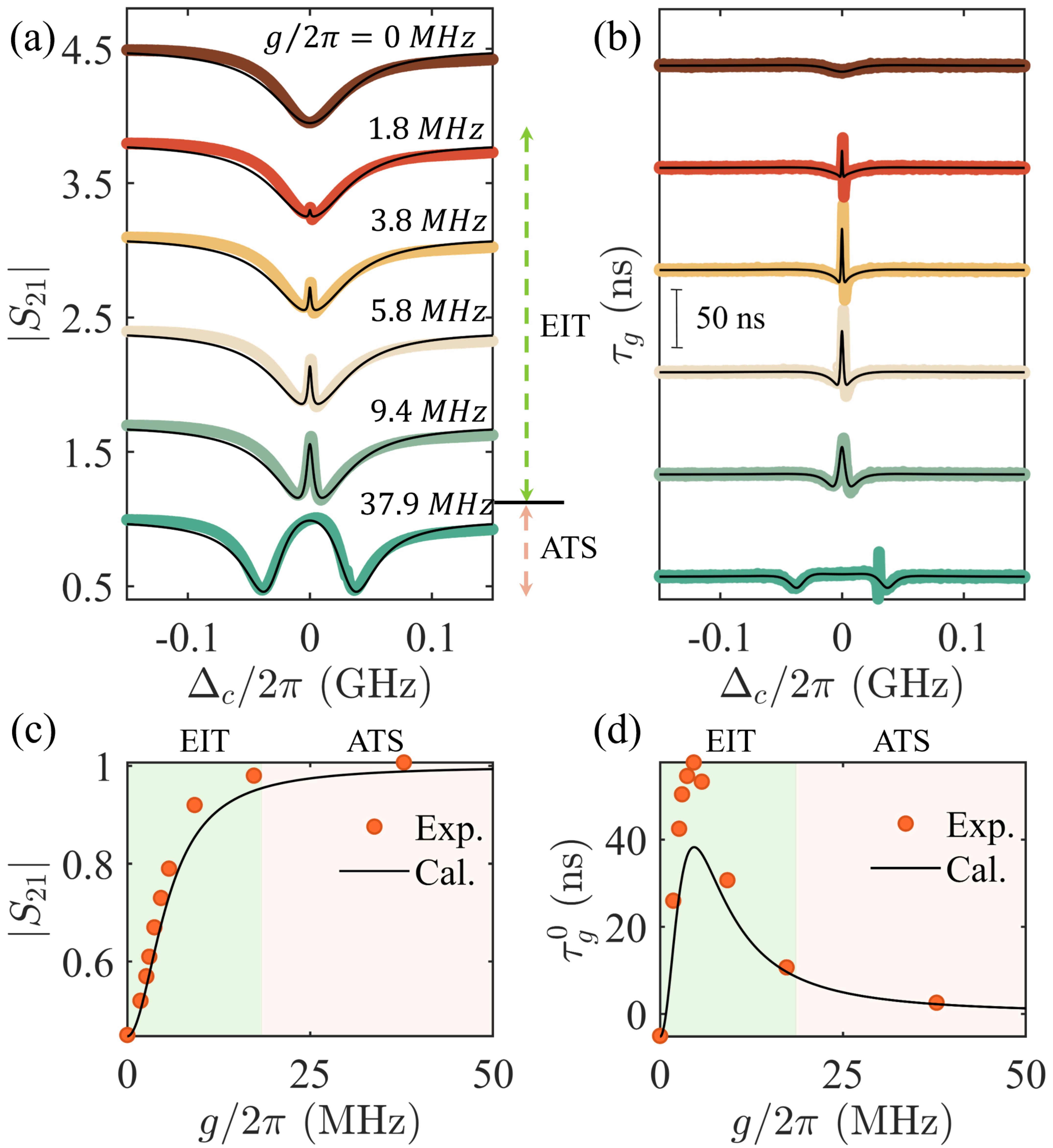}
		\caption{Experimental (a) transmission amplitude $|S_{21}|$ and (b) group delay $\tau_g$ at different coupling strengths $g$ as a function of probing frequency detuning $\Delta_c$ at $\Delta_m=0$. The spectra in panel (a) are shifted upward incrementally by 0.7. The black curves in panels (a) and (b) are the fitting results using Eq. \ref{EIT S21}, and the calculated $\tau_g$ using the fitted $g$, respectively. As $g$ increases, the system transitions from the EIT regime to the ATS regime. The (c) transmission amplitude $|S_{21}|$ and (d) group delay $\tau_g^0$  at $\Delta_m=\Delta_c=0$ as a function of coupling strength $g$, with the green and pink shadows representing the EIT and ATS regimes, respectively. The black curves in panels (c) and (d) are the calculated results using Eqs. \ref{EIT S21} and \ref{group delay as function of g}, respectively.}
		\label{Fig2}
	\end{center}
\end{figure}

Setting $\mu_0 H=221.3$ mT at $\Delta_m=0$, we perform the transmission measurement. The YIG sphere is set at different positions above the DR to change $g$. Figs. \ref{Fig2}(a) and (b) display, respectively, $|S_{21}(\omega)|$ and $\tau_g(\omega)$ measured by sweeping the probing frequency $\omega$. The spectrum at $g=0$ is measured at $H=0$, showing the bare cavity mode with a transmission amplitude of 45\% at $\Delta_c=0$.

The measured $|S_{21}(\omega)|$ spectra (solid marks) in Fig. \ref{Fig2}(a) are fitted (thin curves) by using Eq. \ref{EIT S21} with $g$ as the only fitting parameter. Using the fitted $g$, the calculated $\tau_g(\omega)$ are plotted in Fig. \ref{Fig2}(b) by the thin curves. As $g$ increases $(0<g\le \frac{\kappa_c+\beta_0-\alpha_0}{2})$, the transmission at $\Delta_c=0$ increases, opening up a transparent window superimposed on the broad resonance of the cavity mode, where the hybridized modes have the same frequency but different damping rates. This is the EIT-analog arising from the destructive interference between two microwave pathways: one passing through the YIG sphere and the other bypassing it. When $g$ is large enough $(g>\frac{\kappa_c+\beta_0-\alpha_0}{2})$, the coupled system enters the Autler-Townes splitting (ATS) regime \cite{autler1955stark,peng2014and}. Here, the frequency degeneracy of the eigenmodes is lifted, resulting in two distinct resonances with identical linewidths. This is a typical example of level repulsion caused by coherent magnon-photon coupling \cite{artman1955measurement,huebl2013high,goryachev2014high,tabuchi2014hybridizing,zhang2014strongly,bai2015spin,li2019strong,hou2019strong}.

In Figs. \ref{Fig2}(c) and (d), $|S_{21}|$ and $\tau_g^0$ deduced at $\Delta_c=0$ are plotted as solid marks, respectively. The thin curves represent the calculated data from Eqs. \ref{EIT S21} and \ref{group delay as function of g}, showing reasonable agreement with the observed trends: as $g$ increases, transmission monotonically approaches 100\%, while $\tau_g^0$ rises to a maximum before gradually dropping to zero as the system enters the ATS regime. Deviations in the calculated $\tau_g^0$ may arise from neglecting the geometric details and the mode profile of the DR in the phase analysis.

\textit{Nonreciprocal EIT at $\kappa_m \neq$ \textsc{0}.--}
Now, we configure the device for non-reciprocal EIT by positioning the YIG sphere near the microstrip and adjacent to the DR, with a center-to-center distance of $L \simeq$ 1 cm as shown in Fig. \ref{Fig1}(e). This introduces a small extrinsic damping rate $\kappa_m$ for the magnon mode. In this setup, apart from the direct magnon-photon coupling, the cavity mode also couples with the magnon mode indirectly through traveling waves, leading to dissipative magnon-photon coupling \cite{harder2018level,wang2020dissipative,wang2019nonreciprocity,harder2021coherent} with the rate $\Gamma=\sqrt{\kappa_c \kappa_m}$.

Taking both couplings into account, we derive the transmission parameters \cite{Sup}:
\begin{subequations}
\begin{align}
S_{21}(\omega, \pm H)=\frac{(\omega-\tilde{\omega}_m)(\omega-\tilde{\omega}_c)-g(g\pm 2i\Gamma) }
		{(\omega-\tilde{\omega}_m+i\kappa_m)(\omega-\tilde{\omega}_c+i\kappa_c)-(g^2+\Gamma^2)},\\
S_{12}(\omega, \pm H)=\frac{(\omega-\tilde{\omega}_m)(\omega-\tilde{\omega}_c)-g(g\mp 2i\Gamma) }
		{(\omega-\tilde{\omega}_m+i\kappa_m)(\omega-\tilde{\omega}_c+i\kappa_c)-(g^2+\Gamma^2)},
\end{align}
\label{Nonreciprocal EIT S21}
\end{subequations}
which exhibit signature non-reciprocal relations~\cite{footnote}:
\begin{subequations}
\begin{align}
|S_{21} (\omega, H)| & = |S_{12} (\omega, -H)| \neq |S_{12} (\omega, H)|,\\
\tau_{21} (\omega, H) & = \tau_{12} (\omega, -H) \neq \tau_{12} (\omega, H).
\end{align}
\label{Symmetry}
\end{subequations}
Here, the nonreciprocal group delay $\tau_{21(12)}$ is defined as $\tau_{21(12)}=\partial \angle S_{21(12)}/\partial \omega$. Eq.~\ref{Symmetry} stems from two facts combined (i) the spin precession reverses when the direction of the magnetic field H is reversed, and (ii) the microwave magnetic field reverses when the direction of the traveling wave is reversed. Furthermore, systematical analysis based on Eq. \ref{Nonreciprocal EIT S21} indicates that although $\kappa_m$ introduces non-reciprocity, it reduces the group delay by adding dissipation to the magnon mode \cite{Sup}. Hence, $\kappa_m$ should be carefully engineered at a small level in this scheme, with $\kappa_m < \alpha_0$, to reach a balance between pronounced non-reciprocity and significant group delay.

\begin{figure} [!t]
	\begin{center}
		\includegraphics[width=8.8 cm]{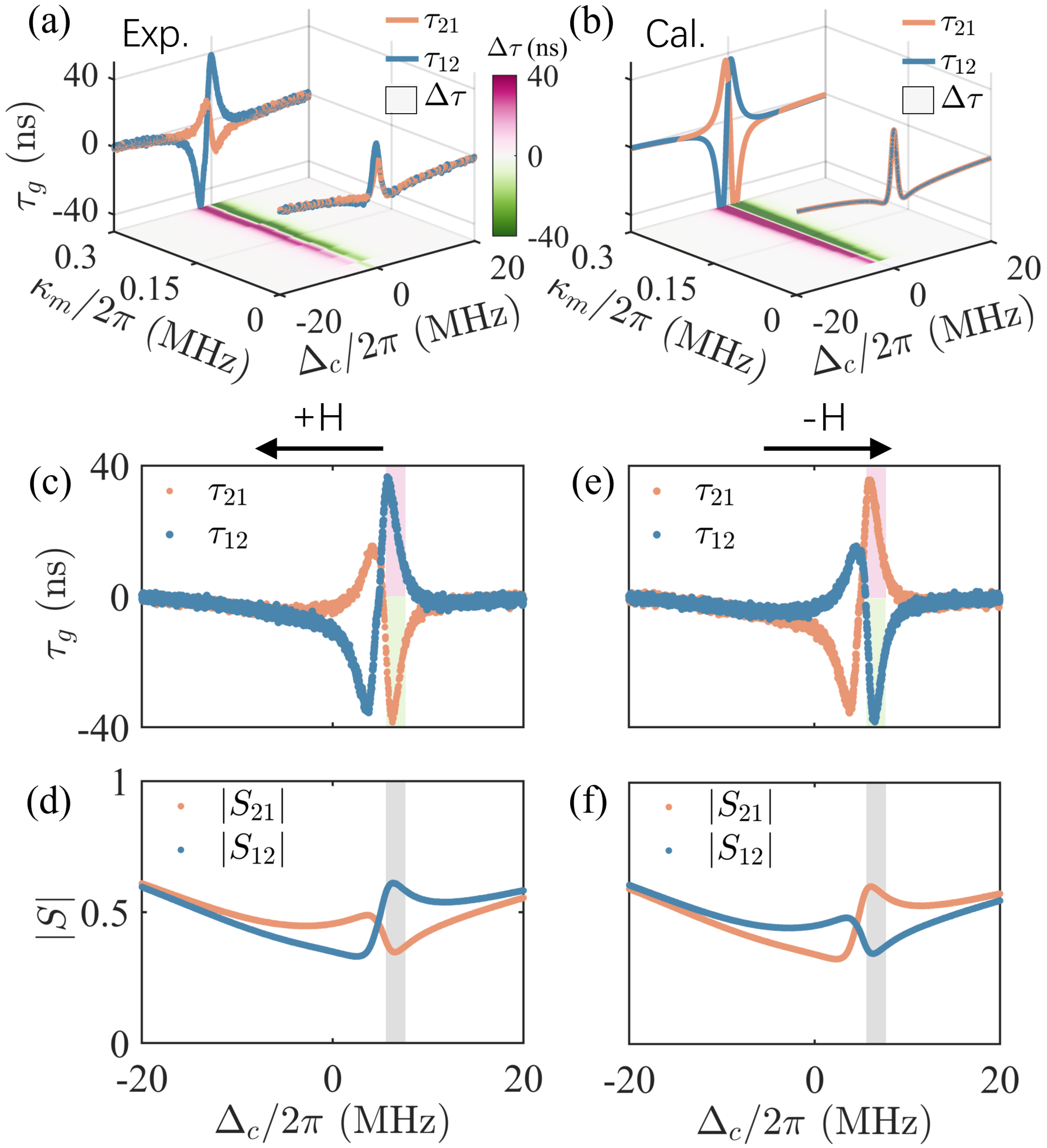}
		\caption{(a),(b) At $g/2\pi=3.1$ MHz and $\Delta_m=0$, the (a) measured and (b) calculated (based on Eq. \ref{Nonreciprocal EIT S21}) $\tau_{21}$ and $\tau_{12}$ as a function of  $\Delta_c$ at different $\kappa_m$. The floors plot $\Delta \tau=\tau_{21}-\tau_{12}$ mappings as functions of $\Delta_c$ and $\kappa_m$. (c)-(f) At the setting of $\kappa_m/2\pi=0.3$ MHz and $\Delta_m/2\pi=4.8$ MHz, the left and right panels compare the results at positive and negative magnetic fields, respectively. The measured (c),(e) group delays ($\tau_{21}$ and $\tau_{12}$) and (d),(f) transmissions ($|S_{21}|$ and $|S_{12}|$) as a function of  $\Delta_c$. The shaded bars denote the range within the full width at half maximum (FWHM) of the frequency spectrum of the Gaussian pulse used in Fig. \ref{Fig4}. }
		\label{Fig3}
	\end{center}
\end{figure}

Setting $g/2\pi=3.1$ MHz at $\Delta_m=0$ as the example, Fig. \ref{Fig3}(a) shows the group delays measured by changing $\kappa_m/2\pi$ from 0 to 0.3 MHz. At $\kappa_m=0$, the result reproduces the reciprocal EIT-analog, manifested as $\tau_{21}(\omega)$ = $\tau_{12}(\omega)$ = $\tau_{g}(\omega)$ that are symmetrically centered at $\Delta_c=0$. At $\kappa_m/2\pi$ = 0.3 MHz, pronounced nonreciprocal windows appear on both sides of $\Delta_c=0$: for $\tau_{21}$, the slow ($\tau_{21} > $ 0) and fast light ($\tau_{21} < $ 0) windows appear at $\Delta_c <$ 0 and $\Delta_c >$ 0, respectively;  while for $\tau_{12}$, the opposite feature is observed. To better display the evolution from EIT-analog to non-reciprocal EIT, we define the nonreciprocal group delay as $\Delta\tau=\tau_{21}-\tau_{12}$, and plot its mappings on the floor of $\Delta_c$ and $\kappa_m$. Clearly, by increasing $\kappa_m$,  non-reciprocal group delays emerge and become increasingly significant. Such an evolution is theoretically well accounted for~\cite{footnote}, as shown by the calculated mapping in Fig. \ref{Fig3}(b).

To verify the signature relations of Eq. \ref{Symmetry}, we set the device at $g/2\pi=3.1$ MHz, $\kappa_m/2\pi=0.3$ MHz and $\Delta_m/2\pi=4.8$ MHz, and we compare its performance by reversing the direction of the $H$-field. Setting the field direction along port 2 to port 1, the measured $\tau_{21(12)}(\omega)$ and $|S_{21(12)}(\omega)|$ are plotted in Figs. \ref{Fig3}(c) and (d), respectively. When the field direction is inverted, the non-reciprocity for both $\tau_{21(12)}(\omega)$ and $|S_{21(12)}(\omega)|$ are directionally reversed, as shown in Figs. \ref{Fig3}(e) and (f), respectively. Remarkably, while the non-reciprocity of the group delay is significant, which enables directional reversal between fast- and slow-light, the bi-directional transmissions retain similar amplitude near 50\%. This makes the cavity magnonics appealing for non-reciprocal time-control of microwave pulses, as we demonstrate below.

\textit{Nonreciprocal time-control of microwave pulses.--}

\begin{figure} [!t]
	\begin{center}
		\includegraphics[width=8.8 cm]{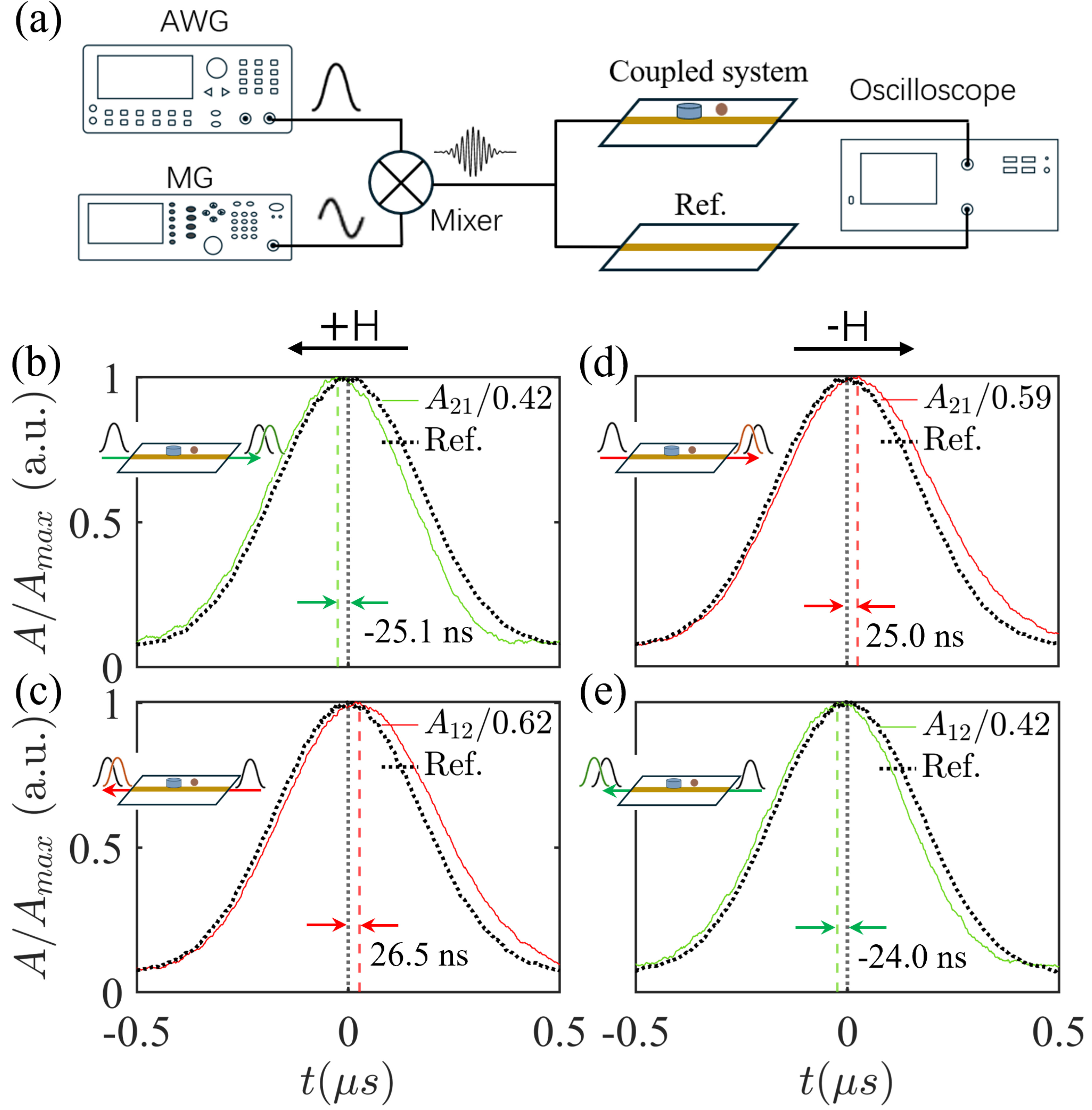}
		\caption{ (a) Schematic of the non-reciprocal pulse control experiment. A Gaussian pulse created by an arbitrary waveform generator (AWG) is mixed with a sinusoidal carrier wave generated by a microwave generator (MG) using a mixer and then split into two. One part passes through the path with the coupled system, while the other passes through a reference path, both measured by an oscilloscope. (b)-(e) The pulse propagation results at (b),(c) positive and (d),(e) negative magnetic fields, while the pulse is sent from (b),(d) port 1 to port 2 and (c),(e) the opposite direction. The black normalized Gaussian pulse envelops denote the pulse through the reference path, while green (red) ones represent the advanced (delayed) pulse through the path with coupled system. $A_{21(12)}$ represents the amplitude of the pulse from port 1(2) to port 2(1). The insets illustrate the schematics of the test setup. }
		\label{Fig4}
	\end{center}
\end{figure}

The pulse control experiment is performed in the time domain by using the setup schematically shown in Fig. \ref{Fig4}(a). The cavity magnonics device is set at the same parameter conditions as that for Figs. \ref{Fig3}(c)-(f). A Gaussian wave pulse with a pulse-width of 400 ns is constructed by an arbitrary waveform generator, and mixed with a sinusoidal carrier sent from a microwave generator operating at $\omega/2\pi=6.2101$ GHz. The pulse is divided into two pathways by a splitter: one passes through the cavity magnonics device, and the other passes through a reference path constructed by identical cables and microstrip \cite{kandic2011asymptotic}, and both are recorded by an oscilloscope. By fitting the measured reference and the delayed (advanced) signal with the Gaussian function, the time delay (advance) $\Delta t$ is deduced from the difference in the maxima of the fitted Gaussian curves.
Physically, $\Delta t$ is the averaged $\tau_g(\omega)$ over the frequency bandwidth of the pulse of about 2 MHz~\cite{Sup}, as depicted by the shaded bars in Figs. \ref{Fig3}(c)-(f). Due to the bandwidth, the transmitted Gaussian pulses will undergo slight distortion compared to the shape of the incident pulse, as will be observed in the measurements below.

Figs. \ref{Fig4}(b) and (c) plot the normalized Gaussian pulse envelop, with the time of the reference pulse's maximum set as time zero. When the pulse transmits from port 1 to port 2 of the device, time advance of $\Delta t$ = -25.1$\pm$0.3 ns is observed, with a transmission of 42\%; when the same pulse transmits from the opposite direction, it shows a 26.5$\pm$0.3 ns delay, with the transmission of 62\%. When the $H$-field direction is inverted, as shown in Figs. \ref{Fig4}(d) and (e), the directional time advance and delay are reversed, in agreement with the signature symmetry displayed in Figs. \ref{Fig3}(c)-(f). On-demand non-reciprocal time control of the microwave pulses is thus achieved.

For clarity, it is necessary to deliberate various forms of the speed of light controlled in this scheme.  The microwave pulse is composed of multiple components with different frequencies and phases, each component travels in the microstrip at the same phase speed $v_p=c/\sqrt{\varepsilon} \approx$ 0.67$c$, where $\varepsilon =2.2$ is the effective permittivity. Estimating the magnon-photon coupling range of $L \approx 1$ cm, the time that microwaves travel at any frequency $\omega$ over this distance is $t_p = L/v_p \approx 50$ ps. Compared with $t_p$, the incident pulses appear to be hugely advanced or delayed by a factor of $|\Delta t|/t_p \ge$ 500. In the same context, using the relation \cite{wang2000gain} $\Delta t=L/v_g-L/v_p$, we can estimate the pulse's group velocity $v_g \approx$ 0.00126$c$ and -0.00133$c$ for the observed slow and fast light, respectively.

The negative $v_g$ appears as if "a negative time" ($\Delta t < -t_p$) was created, causing the pulse peak to exit the device before entering it \cite{Sup}. This intriguing notion has attracted significant interest since the seminal experiment on fast light was conducted using an atomic platform \cite{wang2000gain}. Physically, as noted in the introduction, no information can propagate faster than $c$ or reverse in time. The observed effect can be understood as follows: magnon-photon couplings are initiated as soon as the leading edge of the incident pulse rings up the DR and YIG resonators. Consequently, before the pulse peak even reaches the device, these couplings begin reshaping the pulse by modulating its phase components. This reconstructs an advanced peak leaving the device early, but the actual wavefront always exits the device after entering. Thus, while the group velocity, defined by the pulse peaks, appears negative, the speed of energy flow remains positive and never exceeds $c$.

We conclude by noting that, distinct from conventional pulse techniques for engineering the slow and fast light, this work achieves both effects through magnon-photon couplings. Integrating this discovery with the rapidly evolving field of ultrafast magnonics \cite{igarashi2023optically, dainone2024controlling} is an exciting prospect, as innovations may emerge when various time-domain technologies converge. Moreover, the on-demand non-reciprocal pulse control scheme developed here is unique. This breakthrough, combined with the compact cavity magnonics platform, could unlock previously unimaginable technologies, such as non-reciprocal pulse engineering for microwave communications, asymmetric synapses for neuromorphic networks, and directional pulse regulators for harnessing non-reciprocal entanglement in quantum networks.

This work has been funded by NSERC Discovery Grants and NSERC Discovery Accelerator Supplements (C.-M. H.). We acknowledge CMC Microsystems for providing equipment that facilitated this research, and we thank James Millar and James Dietrich for assistance.

%\bibliographystyle{apsrev4-1}
%\bibliography{v3_nonreciprocal_slow_light}

\begin{thebibliography}{54}%
\makeatletter
\providecommand \@ifxundefined [1]{%
 \@ifx{#1\undefined}
}%
\providecommand \@ifnum [1]{%
 \ifnum #1\expandafter \@firstoftwo
 \else \expandafter \@secondoftwo
 \fi
}%
\providecommand \@ifx [1]{%
 \ifx #1\expandafter \@firstoftwo
 \else \expandafter \@secondoftwo
 \fi
}%
\providecommand \natexlab [1]{#1}%
\providecommand \enquote  [1]{``#1''}%
\providecommand \bibnamefont  [1]{#1}%
\providecommand \bibfnamefont [1]{#1}%
\providecommand \citenamefont [1]{#1}%
\providecommand \href@noop [0]{\@secondoftwo}%
\providecommand \href [0]{\begingroup \@sanitize@url \@href}%
\providecommand \@href[1]{\@@startlink{#1}\@@href}%
\providecommand \@@href[1]{\endgroup#1\@@endlink}%
\providecommand \@sanitize@url [0]{\catcode `\\12\catcode `\$12\catcode
  `\&12\catcode `\#12\catcode `\^12\catcode `\_12\catcode `\%12\relax}%
\providecommand \@@startlink[1]{}%
\providecommand \@@endlink[0]{}%
\providecommand \url  [0]{\begingroup\@sanitize@url \@url }%
\providecommand \@url [1]{\endgroup\@href {#1}{\urlprefix }}%
\providecommand \urlprefix  [0]{URL }%
\providecommand \Eprint [0]{\href }%
\providecommand \doibase [0]{https://doi.org/}%
\providecommand \selectlanguage [0]{\@gobble}%
\providecommand \bibinfo  [0]{\@secondoftwo}%
\providecommand \bibfield  [0]{\@secondoftwo}%
\providecommand \translation [1]{[#1]}%
\providecommand \BibitemOpen [0]{}%
\providecommand \bibitemStop [0]{}%
\providecommand \bibitemNoStop [0]{.\EOS\space}%
\providecommand \EOS [0]{\spacefactor3000\relax}%
\providecommand \BibitemShut  [1]{\csname bibitem#1\endcsname}%
\let\auto@bib@innerbib\@empty
%</preamble>
%ref1
\bibitem [{\citenamefont {Reiskarimian}\ \emph {et~al.}(2016)\citenamefont {Reiskarimian},
 \ and\ \citenamefont {Krishnaswamy}}]{reiskarimian2016magnetic}%
    \BibitemOpen
  \bibfield  {author}
  {\bibinfo {author}
  {\bibfnamefont {N.}~\bibnamefont {Reiskarimian}}, \ and\
  \bibinfo {author}
   {\bibfnamefont {H.}~\bibnamefont {Krishnaswamy}},\ }
  \bibfield {title}
   {\bibinfo {title} {Magnetic-free non-reciprocity based on staggered commutation},\ }\href@noop {}
   {\bibfield  {journal}
  	{\bibinfo  {journal}
  		 {Nat. Commun.}\ }\textbf
  	 {\bibinfo {volume}{7}},\ \bibinfo {pages} {11217} (\bibinfo {year} {2016})}\BibitemShut
  {NoStop}%
%ref2
\bibitem [{\citenamefont {Feng}\ \emph {et~al.}(2011)\citenamefont {Feng}, \citenamefont {Huang},
	\citenamefont {Ayache}, \citenamefont {Huang}, \citenamefont {Xu},\citenamefont {Lu},\citenamefont {Chen},\ and\ \citenamefont  {Fainman}}]{feng2011nonreciprocal}%
\BibitemOpen
\bibfield  {author}
{\bibinfo {author}
	{\bibfnamefont {L.}~\bibnamefont {Feng}},
\bibinfo {author}
	{\bibfnamefont {M.}~\bibnamefont {Ayache}},
\bibinfo {author}
	{\bibfnamefont {J.}~\bibnamefont {Huang}},
\bibinfo {author}
	{\bibfnamefont {Y.-L.}~\bibnamefont {Xu}},
\bibinfo {author}
	{\bibfnamefont {M.-H.}~\bibnamefont {Lu}},
\bibinfo {author}
	{\bibfnamefont {Y.-F.}~\bibnamefont {Chen}},
\bibinfo {author}
	{\bibfnamefont {Y.}~\bibnamefont {Fainman}},
		\ and\
\bibinfo {author}
		{\bibfnamefont {A.}~\bibnamefont {Scherer}},\ }	
\bibfield  {title}
{\bibinfo {title}
	{Nonreciprocal light propagation in a silicon photonic circuit},\ }\href@noop {}
{\bibfield  {journal}
{\bibinfo  {journal} {Science}\ }\textbf {\bibinfo
	{volume} {333}},\ \bibinfo {pages} {729} (\bibinfo {year} {2011})}\BibitemShut
{NoStop}%	
%ref3
\bibitem [{\citenamefont {Shen}\ \emph {et~al.}(2023)\citenamefont {Shen},
	\citenamefont {Zhang}, \citenamefont {Chen}, \citenamefont {Xiao},\citenamefont {Zou},\citenamefont {Guo},\ and\ \citenamefont {Dong}}]{shen2023nonreciprocal}%
\BibitemOpen
\bibfield  {author}
{\bibinfo {author}
	{\bibfnamefont {Z.}~\bibnamefont {Shen}},
\bibinfo {author}
	{\bibfnamefont {Y.-L.}~\bibnamefont {Zhang}},
\bibinfo {author}
	{\bibfnamefont {Y.}~\bibnamefont {Chen}},
\bibinfo {author}
	{\bibfnamefont {Y.-F.}~\bibnamefont {Xiao}},
\bibinfo {author}
	{\bibfnamefont {C.-L.}~\bibnamefont {Zou}},
\bibinfo {author}
	{\bibfnamefont {G.-C.}~\bibnamefont {Guo}},
	\ and\
\bibinfo {author}
	{\bibfnamefont {C.-H.}~\bibnamefont {Dong}},\ }
\bibfield  {title}
{\bibinfo {title}
	{Nonreciprocal frequency conversion and mode routing in a microresonator},\ }\href@noop {}
{\bibfield  {journal}
{\bibinfo  {journal} {Phys. Rev. Lett.}\ }\textbf {\bibinfo
		{volume} {130}},\ \bibinfo {pages} {013601} (\bibinfo {year} {2023})}\BibitemShut
{NoStop}%	
%ref4
\bibitem [{\citenamefont {Rosario}\ \emph {et~al.}(2023)\citenamefont {Rosario},
	\citenamefont {M\"uller}, \citenamefont {Jerger}, \citenamefont {Zanner},\citenamefont {Combes},\citenamefont {Pletyukhov},\citenamefont {Weides},\citenamefont {Stace},\ and\ \citenamefont {Fedorov}}]{rosario2018nonreciprocity}%
\BibitemOpen
\bibfield  {author}
{\bibinfo {author}
	{\bibfnamefont {A. }~\bibnamefont {Rosario Hamann}},
\bibinfo {author}
	{\bibfnamefont {C.}~\bibnamefont {M\"uller}},
\bibinfo {author}
	{\bibfnamefont {M.}~\bibnamefont {Jerger}},
\bibinfo {author}
	{\bibfnamefont {M.}~\bibnamefont {Zanner}},
\bibinfo {author}
	{\bibfnamefont {J.}~\bibnamefont {Combes}},
\bibinfo {author}
	{\bibfnamefont {M.}~\bibnamefont {Pletyukhov}},
\bibinfo {author}
	{\bibfnamefont {M.}~\bibnamefont {Weides}},
\bibinfo {author}
	{\bibfnamefont {T. M.}~\bibnamefont {Stace}},
	\ and\
\bibinfo {author}
	{\bibfnamefont {A. }~\bibnamefont {Fedorov}},\ }
\bibfield  {title}
{\bibinfo {title}
	{Nonreciprocity realized with quantum nonlinearity},\ }\href@noop {}
{\bibfield  {journal}
	{\bibinfo  {journal} {Phys. Rev. Lett.}\ }\textbf {\bibinfo
		{volume} {121}},\ \bibinfo {pages} {123601} (\bibinfo {year} {2018})}\BibitemShut
{NoStop}%	
%ref5
\bibitem [{\citenamefont {Hau}\ \emph {et~al.}(1999)\citenamefont {Hau},
	\citenamefont {Harris}, \citenamefont {Dutton}, \ and\ \citenamefont {Behroozi}}]{hau1999light}%
\BibitemOpen
\bibfield  {author}
{\bibinfo {author}
	{\bibfnamefont {L. V.}~\bibnamefont {Hau}},
\bibinfo {author}
	{\bibfnamefont {S. E.}~\bibnamefont {Harris}},
\bibinfo {author}
	{\bibfnamefont {Z.}~\bibnamefont {Dutton}},
	\ and\
\bibinfo {author}
	{\bibfnamefont {C. H.}~\bibnamefont {Behroozi}},\ }
\bibfield  {title}
{\bibinfo {title}
	{Light speed reduction to 17 metres per second in an ultracold atomic gas},\ }\href@noop {}
{\bibfield  {journal}
	{\bibinfo  {journal} {Nature}\ }\textbf {\bibinfo
		{volume} {397}},\ \bibinfo {pages} {594} (\bibinfo {year} {1999})}\BibitemShut
{NoStop}%
%ref6
\bibitem [{\citenamefont {Bigelow}\ \emph {et~al.}(2003)\citenamefont {Bigelow},
	\citenamefont {Lepeshkin}, \ and\ \citenamefont {Boyd}}]{bigelow2003observation}%
\BibitemOpen
\bibfield  {author}
{\bibinfo {author}
	{\bibfnamefont {M. S.}~\bibnamefont {Bigelow}},
	\bibinfo {author}
	{\bibfnamefont {N. N.}~\bibnamefont {Lepeshkin}},
	\ and\
	\bibinfo {author}
	{\bibfnamefont {R. W.}~\bibnamefont {Boyd}},\ }
\bibfield  {title}
{\bibinfo {title}
	{Observation of ultraslow light propagation in a ruby crystal at room temperature},\ }\href@noop {}
{\bibfield  {journal}
	{\bibinfo  {journal} {Phys. Rev. Lett.}\ }\textbf {\bibinfo
		{volume} {90}},\ \bibinfo {pages} {113903} (\bibinfo {year} {2003})}\BibitemShut
{NoStop}%
%ref7
\bibitem [{\citenamefont {Totsuka}\ \emph {et~al.}(2007)\citenamefont {Totsuka},
	\citenamefont {Kobayashi}, \ and\ \citenamefont {Tomita}}]{totsuka2007slow}%
\BibitemOpen
\bibfield  {author}
{\bibinfo {author}
	{\bibfnamefont {K.}~\bibnamefont {Totsuka}},
	\bibinfo {author}
	{\bibfnamefont {N.}~\bibnamefont {Kobayashi}},
	\ and\
	\bibinfo {author}
	{\bibfnamefont {M. }~\bibnamefont {Tomita}},\ }
\bibfield  {title}
{\bibinfo {title}
	{Slow light in coupled-resonator-induced transparency},\ }\href@noop {}
{\bibfield  {journal}
	{\bibinfo  {journal} {Phys. Rev. Lett.}\ }\textbf {\bibinfo
		{volume} {98}},\ \bibinfo {pages} {213904} (\bibinfo {year} {2007})}\BibitemShut
{NoStop}%
%ref8
\bibitem [{\citenamefont {Amili}\ \emph {et~al.}(2010)\citenamefont {Amili},
	\citenamefont {Miranda}, \citenamefont {Goldfarb}, \citenamefont {Baili},\citenamefont {Beaudoin},\citenamefont {Sagnes},\citenamefont {Bretenaker},\ and\ \citenamefont {Alouini}}]{el2010observation}%
\BibitemOpen
\bibfield  {author}
{\bibinfo {author}
	{\bibfnamefont {A. El}~\bibnamefont {Amili}},
\bibinfo {author}
	{\bibfnamefont {B.-X.}~\bibnamefont {Miranda}},
\bibinfo {author}
	{\bibfnamefont {F.}~\bibnamefont {Goldfarb}},
\bibinfo {author}
	{\bibfnamefont {G.}~\bibnamefont {Baili}},
\bibinfo {author}
	{\bibfnamefont {G.}~\bibnamefont {Beaudoin}},
\bibinfo {author}
	{\bibfnamefont {I.}~\bibnamefont {Sagnes}},
\bibinfo {author}
	{\bibfnamefont {F.}~\bibnamefont {Bretenaker}},
	\ and\
\bibinfo {author}
	{\bibfnamefont {M.}~\bibnamefont {Alouini}},\ }
\bibfield  {title}
{\bibinfo {title}
	{Observation of slow light in the noise spectrum of a vertical external cavity surface-emitting laser},\ }\href@noop {}
{\bibfield  {journal}
	{\bibinfo  {journal} {Phys. Rev. Lett.}\ }\textbf {\bibinfo
		{volume} {105}},\ \bibinfo {pages} {223902} (\bibinfo {year} {2010})}\BibitemShut
{NoStop}%	
%ref9
\bibitem [{\citenamefont {Safavi-Naeini}\ \emph {et~al.}(2011)\citenamefont {Safavi-Naeini},
	\citenamefont {Alegre}, \citenamefont {Chan}, \citenamefont {Eichenﬁeld},\citenamefont {Winger},\citenamefont {Lin},\citenamefont {Hill},\citenamefont {Chang},\ and\ \citenamefont {Painter}}]{safavi2011electromagnetically}%
\BibitemOpen
\bibfield  {author}
{\bibinfo {author}
	{\bibfnamefont {A. H.}~\bibnamefont {Safavi-Naeini}},
\bibinfo {author}
	{\bibfnamefont {T. M.}~\bibnamefont {Alegre}},
\bibinfo {author}
	{\bibfnamefont {J.}~\bibnamefont {Chan}},
\bibinfo {author}
	{\bibfnamefont {M.}~\bibnamefont {Eichenﬁeld}},
\bibinfo {author}
	{\bibfnamefont {M.}~\bibnamefont {Winger}},
\bibinfo {author}
	{\bibfnamefont {Q.}~\bibnamefont {Lin}},
\bibinfo {author}
	{\bibfnamefont {J. T.}~\bibnamefont {Hill}},
\bibinfo {author}
	{\bibfnamefont {D. E.}~\bibnamefont {Chang}},
	\ and\
\bibinfo {author}
	{\bibfnamefont {O. }~\bibnamefont {Painter}},\ }
\bibfield  {title}
{\bibinfo {title}
	{Electromagnetically induced transparency and slow light with optomechanics},\ }\href@noop {}
{\bibfield  {journal}
	{\bibinfo  {journal} {Nature}\ }\textbf {\bibinfo
		{volume} {472}},\ \bibinfo {pages} {69} (\bibinfo {year} {2011})}\BibitemShut
{NoStop}%
%ref10
\bibitem [{\citenamefont {Heeg}\ \emph {et~al.}(2015)\citenamefont {Heeg},\citenamefont    {Haber},\citenamefont {Schumacher},
	\citenamefont {Bocklage}, \citenamefont {Wille}, \citenamefont {Schulze},\citenamefont {Loetzsch},\citenamefont {Uschmann},\citenamefont {Paulus},\citenamefont {R\"uﬀer}\emph {et~al.}}]{heeg2015tunable}%
\BibitemOpen
\bibfield  {author}
{\bibinfo {author}
	{\bibfnamefont {K. P.}~\bibnamefont {Heeg}},
\bibinfo {author}
	{\bibfnamefont {J.}~\bibnamefont {Haber}},
\bibinfo {author}
	{\bibfnamefont {D.}~\bibnamefont {Schumacher}},
\bibinfo {author}
	{\bibfnamefont {L.}~\bibnamefont {Bocklage}},
\bibinfo {author}
	{\bibfnamefont {H.-C.}~\bibnamefont {Wille}},
\bibinfo {author}
	{\bibfnamefont {K. S.}~\bibnamefont {Schulze}},
\bibinfo {author}
	{\bibfnamefont {R.}~\bibnamefont {Loetzsch}},
\bibinfo {author}
	{\bibfnamefont {I.}~\bibnamefont {Uschmann}},
\bibinfo {author}
	{\bibfnamefont {G. G.}~\bibnamefont {Paulus}},
\bibinfo {author}
	{\bibfnamefont {R.}~\bibnamefont {R\"uﬀer}}, \emph {et~al.},\ }
\bibfield  {title}
{\bibinfo {title}
	{Tunable subluminal propagation of narrow-band x-ray pulses},\ }\href@noop {}
{\bibfield  {journal}
	{\bibinfo  {journal} {Phys. Rev. Lett.}\ }\textbf {\bibinfo
		{volume} {114}},\ \bibinfo {pages} {203601} (\bibinfo {year} {2015})}\BibitemShut
{NoStop}%
%ref11
\bibitem [{\citenamefont {Wang}\ \emph {et~al.}(2000)\citenamefont {Wang},
	\citenamefont {Kuzmich}, \ and\ \citenamefont {Dogariu}}]{wang2000gain}%
\BibitemOpen
\bibfield  {author}
{\bibinfo {author}
	{\bibfnamefont {L. J.}~\bibnamefont {Wang}},
	\bibinfo {author}
	{\bibfnamefont {A.}~\bibnamefont {Kuzmich}},
	\ and\
	\bibinfo {author}
	{\bibfnamefont {A.}~\bibnamefont {Dogariu}},\ }
\bibfield  {title}
{\bibinfo {title}
	{Gain-assisted superluminal light propagation},\ }\href@noop {}
{\bibfield  {journal}
	{\bibinfo  {journal} {Nature}\ }\textbf {\bibinfo
		{volume} {406}},\ \bibinfo {pages} {277} (\bibinfo {year} {2000})}\BibitemShut
{NoStop}%
%ref12
\bibitem [{\citenamefont {Bigelow}\ \emph {et~al.}(2003)\citenamefont {Bigelow},
	\citenamefont {Lepeshkin}, \ and\ \citenamefont {Boyd}}]{bigelow2003superluminal}%
\BibitemOpen
\bibfield  {author}
{\bibinfo {author}
	{\bibfnamefont {M. S.}~\bibnamefont {Bigelow}},
	\bibinfo {author}
	{\bibfnamefont {N. N.}~\bibnamefont {Lepeshkin}},
	\ and\
	\bibinfo {author}
	{\bibfnamefont {R. W.}~\bibnamefont {Boyd}},\ }
\bibfield  {title}
{\bibinfo {title}
	{Superluminal and slow light propagation in a room-temperature solid},\ }\href@noop {}
{\bibfield  {journal}
	{\bibinfo  {journal} {Science}\ }\textbf {\bibinfo
		{volume} {301}},\ \bibinfo {pages} {200} (\bibinfo {year} {2003})}\BibitemShut
{NoStop}%
%ref13
\bibitem [{\citenamefont {Goyon}\ \emph {et~al.}(2021)\citenamefont {Goyon},
	\citenamefont {Edwards}, \citenamefont {Chapman}, \citenamefont {Divol},\citenamefont {Lemos},\citenamefont {Williams},\citenamefont {Mariscal},\citenamefont {Turnbull},\citenamefont {Hansen},\ and\ \citenamefont {Michel}}]{goyon2021slow}%
\BibitemOpen
\bibfield  {author}
{\bibinfo {author}
	{\bibfnamefont {G.}~\bibnamefont {Goyon}},
\bibinfo {author}
	{\bibfnamefont {M.}~\bibnamefont {Edwards}},
\bibinfo {author}
	{\bibfnamefont {T.}~\bibnamefont {Chapman}},
\bibinfo {author}
	{\bibfnamefont {L.}~\bibnamefont {Divol}},
\bibinfo {author}
	{\bibfnamefont {N.}~\bibnamefont {Lemos}},
\bibinfo {author}
	{\bibfnamefont {G.}~\bibnamefont {Williams}},
\bibinfo {author}
	{\bibfnamefont {D.}~\bibnamefont {Mariscal}},
\bibinfo {author}
	{\bibfnamefont {D.}~\bibnamefont {Turnbull}},
\bibinfo {author}
	{\bibfnamefont {A.}~\bibnamefont {Hansen}},
	\ and\
	\bibinfo {author}
	{\bibfnamefont {P.}~\bibnamefont {Michel}},\ }
\bibfield  {title}
{\bibinfo {title}
	{Slow and fast light in plasma using optical wave mixing},\ }\href@noop {}
{\bibfield  {journal}
	{\bibinfo  {journal} {Phys. Rev. Lett.}\ }\textbf {\bibinfo
		{volume} {126}},\ \bibinfo {pages} {205001} (\bibinfo {year} {2021})}\BibitemShut
{NoStop}%	
%ref14
\bibitem [{\citenamefont {Phillips}\ \emph {et~al.}(2001)\citenamefont {Phillips},\citenamefont {Fleischhauer},\citenamefont {Mair},\citenamefont {Walsworth}, \ and\ \citenamefont {Lukin}}]{phillips2001storage}%
\BibitemOpen
\bibfield  {author}
{\bibinfo {author}
	{\bibfnamefont {D. F.}~\bibnamefont {Phillips}},
\bibinfo {author}
	{\bibfnamefont {A.}~\bibnamefont {Fleischhauer}},
\bibinfo {author}
	{\bibfnamefont {A.}~\bibnamefont {Mair}},
	\bibinfo {author}
	{\bibfnamefont {R. L.}~\bibnamefont {Walsworth}},
	\ and\
\bibinfo {author}
	{\bibfnamefont {M. D. }~\bibnamefont {Lukin}},\ }
\bibfield  {title}
{\bibinfo {title}
	{Storage of light in atomic vapor},\ }\href@noop {}
{\bibfield  {journal}
	{\bibinfo  {journal} {Phys. Rev. Lett.}\ }\textbf {\bibinfo
		{volume} {86}},\ \bibinfo {pages} {783} (\bibinfo {year} {2001})}\BibitemShut
{NoStop}%
%ref15
\bibitem [{\citenamefont {Liu}\ \emph {et~al.}(2001)\citenamefont {Liu},
	\citenamefont {Dutton},\citenamefont {Behroozi}, \ and\ \citenamefont {Hau}}]{liu2001observation}%
\BibitemOpen
\bibfield  {author}
{\bibinfo {author}
	{\bibfnamefont {C. }~\bibnamefont {Liu}},
\bibinfo {author}
	{\bibfnamefont {Z.}~\bibnamefont {Dutton}},
\bibinfo {author}
	{\bibfnamefont {C. H.}~\bibnamefont {Behroozi}},
	\ and\
\bibinfo {author}
	{\bibfnamefont {L. V.}~\bibnamefont {Hau}},\ }
\bibfield  {title}
{\bibinfo {title}
	{Observation of coherent optical information storage in an atomic medium using halted light pulses},\ }\href@noop {}
{\bibfield  {journal}
	{\bibinfo  {journal} {Nature}\ }\textbf {\bibinfo
		{volume} {409}},\ \bibinfo {pages} {490} (\bibinfo {year} {2001})}\BibitemShut
{NoStop}%
%ref16
\bibitem [{\citenamefont {Heinze}\ \emph {et~al.}(2013)\citenamefont {Heinze},
	\citenamefont {Hubrich}, \ and\ \citenamefont {Halfmann}}]{heinze2013stopped}%
\BibitemOpen
\bibfield  {author}
{\bibinfo {author}
	{\bibfnamefont {G.}~\bibnamefont {Heinze}},
\bibinfo {author}
	{\bibfnamefont {C.}~\bibnamefont {Hubrich}},
	\ and\
\bibinfo {author}
	{\bibfnamefont {T.}~\bibnamefont {Halfmann}},\ }
\bibfield  {title}
{\bibinfo {title}
	{Stopped light and image storage by electromagnetically induced transparency up to the regime of one minute},\ }\href@noop {}
{\bibfield  {journal}
	{\bibinfo  {journal} {Phys. Rev. Lett.}\ }\textbf {\bibinfo
		{volume} {111}},\ \bibinfo {pages} {033601} (\bibinfo {year} {2013})}\BibitemShut
{NoStop}%
%ref17
\bibitem [{\citenamefont {Lake}\ \emph {et~al.}(2021)\citenamefont {Lake},\citenamefont {Mitchell},\citenamefont {Sukachev},
	\ and\ \citenamefont {Barclay}}]{lake2021processing}%
\BibitemOpen
\bibfield  {author}
{\bibinfo {author}
	{\bibfnamefont {D. P.}~\bibnamefont {Lake}},
\bibinfo {author}
	{\bibfnamefont {M.}~\bibnamefont {Mitchell}},
\bibinfo {author}
	{\bibfnamefont {D. D.}~\bibnamefont {Sukachev}},
	 \ and\
\bibinfo {author}
	{\bibfnamefont {P. E.}~\bibnamefont {Barclay}},\ }
\bibfield {title}
{\bibinfo {title} {Processing light with an optically tunable mechanical memory},\ }\href@noop {}
{\bibfield  {journal}
	{\bibinfo  {journal}
		{Nat. Commun.}\ }\textbf
	{\bibinfo {volume}{12}},\ \bibinfo {pages} {663} (\bibinfo {year} {2021})}\BibitemShut
{NoStop}%
%ref18
\bibitem [{\citenamefont {Franke-Arnold}\ \emph {et~al.}(2011)\citenamefont {Franke-Arnold},
	\citenamefont {Gibson},\citenamefont {Boyd}, \ and\ \citenamefont {Padgett}}]{franke2011rotary}%
\BibitemOpen
\bibfield  {author}
{\bibinfo {author}
	{\bibfnamefont {S.}~\bibnamefont {Franke-Arnold}},
\bibinfo {author}
	{\bibfnamefont {G.}~\bibnamefont {Gibson}},
\bibinfo {author}
	{\bibfnamefont {R. W.}~\bibnamefont {Boyd}},
	\ and\
\bibinfo {author}
	{\bibfnamefont {M. J.}~\bibnamefont {Padgett}},\ }
\bibfield  {title}
{\bibinfo {title}
	{Rotary photon drag enhanced by a slow-light medium},\ }\href@noop {}
{\bibfield  {journal}
	{\bibinfo  {journal} {Science}\ }\textbf {\bibinfo
		{volume} {333}},\ \bibinfo {pages} {65} (\bibinfo {year} {2011})}\BibitemShut
{NoStop}%
%ref19
\bibitem [{\citenamefont {Huet}\ \emph {et~al.}(2016)\citenamefont {Huet},
	\citenamefont {Rasoloniaina}, \citenamefont {Guillem\'e}, \citenamefont {Rochard}, \citenamefont {F\'eron},\citenamefont {Mortier},\citenamefont {Levenson},\citenamefont {Bencheikh},\citenamefont {Yacomotti},\ and\ \citenamefont {Dumeige}}]{huet2016millisecond}%
\BibitemOpen
\bibfield  {author}
{\bibinfo {author}
	{\bibfnamefont {V.}~\bibnamefont {Huet}},
\bibinfo {author}
	{\bibfnamefont {A.}~\bibnamefont {Rasoloniaina}},
\bibinfo {author}
	{\bibfnamefont {P.}~\bibnamefont {Guillem{\'e}}},
\bibinfo {author}
	{\bibfnamefont {P.}~\bibnamefont {Rochard}},
\bibinfo {author}
	{\bibfnamefont {P.}~\bibnamefont {F{\'e}ron}},
\bibinfo {author}
	{\bibfnamefont {M.}~\bibnamefont {Mortier}},
\bibinfo {author}
	{\bibfnamefont {A.}~\bibnamefont {Levenson}},
\bibinfo {author}
	{\bibfnamefont {K.}~\bibnamefont {Bencheikh}},
\bibinfo {author}
	{\bibfnamefont {A.}~\bibnamefont {Yacomotti}},
	\ and\
\bibinfo {author}
	{\bibfnamefont {Y.}~\bibnamefont {Dumeige}},\ }
\bibfield  {title}
{\bibinfo {title}
	{Millisecond photon lifetime in a slow-light microcavity},\ }\href@noop {}
{\bibfield  {journal}
	{\bibinfo  {journal} {Phys. Rev. Lett.}\ }\textbf {\bibinfo
		{volume} {116}},\ \bibinfo {pages} {133902} (\bibinfo {year} {2016})}\BibitemShut
{NoStop}%	
%ref20
\bibitem [{\citenamefont {Xu}\ \emph {et~al.}(2024)\citenamefont {Xu},\citenamefont    {Zhong},\citenamefont {Zhuang},
	\citenamefont {Qian}, \citenamefont {Jiang}, \citenamefont {Pishehvar},\citenamefont {Han},\citenamefont {Jin},\citenamefont {Jornet},\citenamefont {Zhen}\emph {et~al.}}]{xu2024slow}%
\BibitemOpen
\bibfield  {author}
{\bibinfo {author}
	{\bibfnamefont {J.}~\bibnamefont {Xu}},
\bibinfo {author}
	{\bibfnamefont {C.}~\bibnamefont {Zhong}},
\bibinfo {author}
	{\bibfnamefont {S.}~\bibnamefont {Zhuang}},
\bibinfo {author}
	{\bibfnamefont {C.}~\bibnamefont {Qian}},
\bibinfo {author}
	{\bibfnamefont {Y.}~\bibnamefont {Jiang}},
\bibinfo {author}
	{\bibfnamefont {A.}~\bibnamefont {Pishehvar}},
\bibinfo {author}
	{\bibfnamefont {X.}~\bibnamefont {Han}},
\bibinfo {author}
	{\bibfnamefont {D.}~\bibnamefont {Jin}},
\bibinfo {author}
	{\bibfnamefont {J. M.}~\bibnamefont {Jornet}},
\bibinfo {author}
	{\bibfnamefont {B.}~\bibnamefont {Zhen}}, \emph {et~al.},\ }
\bibfield  {title}
{\bibinfo {title}
	{Slow-wave hybrid magnonics},\ }\href@noop {}
{\bibfield  {journal}
	{\bibinfo  {journal} {Phys. Rev. Lett.}\ }\textbf {\bibinfo
		{volume} {132}},\ \bibinfo {pages} {116701} (\bibinfo {year} {2024})}\BibitemShut
{NoStop}%
%ref21
\bibitem [{\citenamefont {Boller}\ \emph {et~al.}(1991)\citenamefont {Boller},
	\citenamefont {Imamo\u{g}lu}, \ and\ \citenamefont {Harris}}]{boller1991observation}%
\BibitemOpen
\bibfield  {author}
{\bibinfo {author}
	{\bibfnamefont {K.-J.}~\bibnamefont {Boller}},
\bibinfo {author}
	{\bibfnamefont {A.}~\bibnamefont {Imamo\u{g}lu}},
	\ and\
\bibinfo {author}
	{\bibfnamefont {S. E.}~\bibnamefont {Harris}},\ }
\bibfield  {title}
{\bibinfo {title}
	{Observation of electromagnetically induced transparency},\ }\href@noop {}
{\bibfield  {journal}
	{\bibinfo  {journal} {Phys. Rev. Lett.}\ }\textbf {\bibinfo
		{volume} {66}},\ \bibinfo {pages} {2593} (\bibinfo {year} {1991})}\BibitemShut
{NoStop}%
%ref22
\bibitem [{\citenamefont {Fleischhauer}\ \emph {et~al.}(2005)\citenamefont {Fleischhauer},
	\citenamefont {Imamoglu}, \ and\ \citenamefont {Marangos}}]{fleischhauer2005electromagnetically}%
\BibitemOpen
\bibfield  {author}
{\bibinfo {author}
	{\bibfnamefont {M.}~\bibnamefont {Fleischhauer}},
\bibinfo {author}
	{\bibfnamefont {A.}~\bibnamefont {Imamoglu}},
	\ and\
\bibinfo {author}
	{\bibfnamefont {J. P.}~\bibnamefont {Marangos}},\ }
\bibfield  {title}
{\bibinfo {title}
	{Electromagnetically induced transparency: Optics in coherent media},\ }\href@noop {}
{\bibfield  {journal}
	{\bibinfo  {journal} {Rev. Mod. Phys.}\ }\textbf {\bibinfo
		{volume} {77}},\ \bibinfo {pages} {633} (\bibinfo {year} {2005})}\BibitemShut
{NoStop}%
%ref23
\bibitem [{\citenamefont {Smith}\ \emph {et~al.}(2004)\citenamefont {Smith},\citenamefont {Chang},\citenamefont {Fuller},\citenamefont {Rosenberger}, \ and\ \citenamefont {Boyd}}]{smith2004coupled}%
\BibitemOpen
\bibfield  {author}
{\bibinfo {author}
	{\bibfnamefont {D. D.}~\bibnamefont {Smith}},
\bibinfo {author}
	{\bibfnamefont {H.}~\bibnamefont {Chang}},
\bibinfo {author}
	{\bibfnamefont {K. A.}~\bibnamefont {Fuller}},
\bibinfo {author}
	{\bibfnamefont {A.}~\bibnamefont {Rosenberger}},
	\ and\
\bibinfo {author}
	{\bibfnamefont {R. W.}~\bibnamefont {Boyd}},\ }
\bibfield  {title}
{\bibinfo {title}
	{Coupled-resonator-induced transparency},\ }\href@noop {}
{\bibfield  {journal}
	{\bibinfo  {journal} {Phys. Rev. A}\ }\textbf {\bibinfo
		{volume} {69}},\ \bibinfo {pages} {063804} (\bibinfo {year} {2004})}\BibitemShut
{NoStop}%
%ref24
\bibitem [{\citenamefont {Liu}\ \emph {et~al.}(2017)\citenamefont {Liu},
	\citenamefont {Li}, \ and\ \citenamefont {Xiao}}]{liu2017electromagnetically}%
\BibitemOpen
\bibfield  {author}
{\bibinfo {author}
	{\bibfnamefont {Y.-C.}~\bibnamefont {Liu}},
\bibinfo {author}
	{\bibfnamefont {B.-B. }~\bibnamefont {Li}},
	\ and\
\bibinfo {author}
	{\bibfnamefont {Y.-F. }~\bibnamefont {Xiao}},\ }
\bibfield  {title}
{\bibinfo {title}
	{Electromagnetically induced transparency in optical microcavities},\ }\href@noop {}
{\bibfield  {journal}
	{\bibinfo  {journal} {Nanophotonics}\ }\textbf {\bibinfo
		{volume} {6}},\ \bibinfo {pages} {789} (\bibinfo {year} {2017})}\BibitemShut
{NoStop}%
%ref25
\bibitem [{\citenamefont {Shen}\ \emph {et~al.}(2016)\citenamefont {Shen},
	\citenamefont {Zhang}, \citenamefont {Chen}, \citenamefont {Zou},\citenamefont {Xiao},\citenamefont {Zou},\citenamefont {Sun},\citenamefont {Guo},\ and\ \citenamefont {Dong}}]{shen2016experimental}%
\BibitemOpen
\bibfield  {author}
{\bibinfo {author}
	{\bibfnamefont {Z.}~\bibnamefont {Shen}},
\bibinfo {author}
	{\bibfnamefont {Y.-L.}~\bibnamefont {Zhang}},
\bibinfo {author}
	{\bibfnamefont {Y.}~\bibnamefont {Chen}},
\bibinfo {author}
	{\bibfnamefont {C.-L.}~\bibnamefont {Zou}},
\bibinfo {author}
	{\bibfnamefont {Y.-F.}~\bibnamefont {Xiao}},
\bibinfo {author}
	{\bibfnamefont {X.-B.}~\bibnamefont {Zou}},
\bibinfo {author}
	{\bibfnamefont {F.-W.}~\bibnamefont {Sun}},
\bibinfo {author}
	{\bibfnamefont {G.-C.}~\bibnamefont {Guo}},
	\ and\
\bibinfo {author}
	{\bibfnamefont {C.-H.}~\bibnamefont {Dong}},\ }
\bibfield  {title}
{\bibinfo {title}
	{Experimental realization of optomechanically induced non-reciprocity},\ }\href@noop {}
{\bibfield  {journal}
	{\bibinfo  {journal} {Nat. Photonics}\ }\textbf {\bibinfo
		{volume} {10}},\ \bibinfo {pages} {657} (\bibinfo {year} {2016})}\BibitemShut
{NoStop}%	
%ref26
\bibitem [{\citenamefont {Dong}\ \emph {et~al.}(2015)\citenamefont {Dong}, \citenamefont {Shen}, \citenamefont {Zou},\citenamefont {Zhang},\citenamefont {Fu},\ and\ \citenamefont  {Guo}}]{dong2015brillouin}%
\BibitemOpen
\bibfield  {author}
{\bibinfo {author}
	{\bibfnamefont {C.-H. }~\bibnamefont {Dong}},
\bibinfo {author}
	{\bibfnamefont {Z.}~\bibnamefont {Shen}},
\bibinfo {author}
	{\bibfnamefont {C.-L.}~\bibnamefont {Zou}},
\bibinfo {author}
	{\bibfnamefont {Y.-L.}~\bibnamefont {Zhang}},
\bibinfo {author}
	{\bibfnamefont {W.}~\bibnamefont {Fu}},
	\ and\
\bibinfo {author}
	{\bibfnamefont {G.-C.}~\bibnamefont {Guo}},\ }
\bibfield  {title}
{\bibinfo {title}
	{Brillouin-scattering-induced transparency and non-reciprocal light storage},\ }\href@noop {}
{\bibfield  {journal}
	{\bibinfo  {journal} {Nat. Commun.}\ }\textbf {\bibinfo
		{volume} {6}},\ \bibinfo {pages} {6193} (\bibinfo {year} {2015})}\BibitemShut
{NoStop}%	
%ref27
\bibitem [{\citenamefont {Rameshti}\ \emph {et~al.}(2022)\citenamefont {Rameshti},
	\citenamefont {Kusminskiy}, \citenamefont {Haigh}, \citenamefont {Usami},\citenamefont {Lachance-Quirion},\citenamefont {Nakamura},\citenamefont {Hu},\citenamefont {Tang},\citenamefont {Bauer},\ and\ \citenamefont {Blanter}}]{rameshti2022cavity}%
\BibitemOpen
\bibfield  {author}
{\bibinfo {author}
	{\bibfnamefont {B. Z.}~\bibnamefont {Rameshti}},
\bibinfo {author}
	{\bibfnamefont {S. V. }~\bibnamefont {Kusminskiy}},
\bibinfo {author}
	{\bibfnamefont {J. A.}~\bibnamefont {Haigh}},
\bibinfo {author}
	{\bibfnamefont {K.}~\bibnamefont {Usami}},
\bibinfo {author}
	{\bibfnamefont {D.}~\bibnamefont {Lachance-Quirion}},
\bibinfo {author}
	{\bibfnamefont {Y.}~\bibnamefont {Nakamura}},
\bibinfo {author}
	{\bibfnamefont {C.-M.}~\bibnamefont {Hu}},
\bibinfo {author}
	{\bibfnamefont {H. X.}~\bibnamefont {Tang}},
\bibinfo {author}
	{\bibfnamefont {G. E.}~\bibnamefont {Bauer}},
	\ and\
\bibinfo {author}
	{\bibfnamefont {Y. M.}~\bibnamefont {Blanter}},\ }
\bibfield  {title}
{\bibinfo {title}
	{Cavity magnonics},\ }\href@noop {}
{\bibfield  {journal}
	{\bibinfo  {journal} {Phys. Rep.}\ }\textbf {\bibinfo
		{volume} {979}},\ \bibinfo {pages} {1} (\bibinfo {year} {2022})}\BibitemShut
{NoStop}%	
%ref28
\bibitem [{\citenamefont {Zhang}\ \emph {et~al.}(2014)\citenamefont {Zhang},\citenamefont {Zou},
	\citenamefont {Jiang}, \ and\ \citenamefont {Tang}}]{zhang2014strongly}%
\BibitemOpen
\bibfield  {author}
{\bibinfo {author}
	{\bibfnamefont {X.}~\bibnamefont {Zhang}},
\bibinfo {author}
	{\bibfnamefont {C.-L.}~\bibnamefont {Zou}},
\bibinfo {author}
	{\bibfnamefont {L.}~\bibnamefont {Jiang}},
	\ and\
\bibinfo {author}
	{\bibfnamefont {H. X.}~\bibnamefont {Tang}},\ }
\bibfield  {title}
{\bibinfo {title}
	{Strongly coupled magnons and cavity microwave photons},\ }\href@noop {}
{\bibfield  {journal}
	{\bibinfo  {journal} {Phys. Rev. Lett.}\ }\textbf {\bibinfo
		{volume} {113}},\ \bibinfo {pages} {156401} (\bibinfo {year} {2014})}\BibitemShut
{NoStop}%
%ref29
\bibitem [{\citenamefont {Liu}\ \emph {et~al.}(2019)\citenamefont {Liu},
	\citenamefont {Xiong}, \ and\ \citenamefont {Wu}}]{liu2019room}%
\BibitemOpen
\bibfield  {author}
{\bibinfo {author}
	{\bibfnamefont {Z.-X.}~\bibnamefont {Liu}},
	\bibinfo {author}
	{\bibfnamefont {H.}~\bibnamefont {Xiong}},
	\ and\
	\bibinfo {author}
	{\bibfnamefont {Y.}~\bibnamefont {Wu}},\ }
\bibfield  {title}
{\bibinfo {title}
	{Room-temperature slow light in a coupled cavity magnon-photon system},\ }\href@noop {}
{\bibfield  {journal}
	{\bibinfo  {journal} {IEEE Access}\ }\textbf {\bibinfo
		{volume} {7}},\ \bibinfo {pages} {57047} (\bibinfo {year} {2019})}\BibitemShut
{NoStop}%
%ref30
\bibitem [{\citenamefont {Zhao}\ \emph {et~al.}(2021)\citenamefont {Zhao},\citenamefont {Wu},\citenamefont {Li},\citenamefont {Liu},\citenamefont {Nori},\citenamefont {Liu}, \ and\ \citenamefont {Du}}]{zhao2021phase}%
\BibitemOpen
\bibfield  {author}
{\bibinfo {author}
	{\bibfnamefont {J.}~\bibnamefont {Zhao}},
\bibinfo {author}
	{\bibfnamefont {L.}~\bibnamefont {Wu}},
\bibinfo {author}
	{\bibfnamefont {T.}~\bibnamefont {Li}},
\bibinfo {author}
		{\bibfnamefont {Y.-X.}~\bibnamefont {Liu}},
\bibinfo {author}
		{\bibfnamefont {F.}~\bibnamefont {Nori}},
\bibinfo {author}
	{\bibfnamefont {Y.}~\bibnamefont {Liu}},
	\ and\
\bibinfo {author}
	{\bibfnamefont {J.}~\bibnamefont {Du}},\ }
\bibfield  {title}
{\bibinfo {title}
	{Phase-controlled pathway interferences and switchable fast-slow light in a cavity-magnon polariton system},\ }\href@noop {}
{\bibfield  {journal}
	{\bibinfo  {journal} {Phys. Rev. Appl.}\ }\textbf {\bibinfo
		{volume} {15}},\ \bibinfo {pages} {024056} (\bibinfo {year} {2021})}\BibitemShut
{NoStop}%
%ref31
\bibitem [{\citenamefont {Xiong}\ \emph {et~al.}(2022)\citenamefont {Xiong},\citenamefont    {Inman},\citenamefont {Li},
	\citenamefont {Xie}, \citenamefont {Bidthanapally}, \citenamefont {Sklenar},\citenamefont {Li},\citenamefont {Louis},\citenamefont {Tyberkevych},\citenamefont {Qu}\emph {et~al.}}]{xiong2022tunable}%
\BibitemOpen
\bibfield  {author}
{\bibinfo {author}
	{\bibfnamefont {Y.}~\bibnamefont {Xiong}},
\bibinfo {author}
	{\bibfnamefont {J.}~\bibnamefont {Inman}},
\bibinfo {author}
	{\bibfnamefont {Z.}~\bibnamefont {Li}},
\bibinfo {author}
	{\bibfnamefont {K.}~\bibnamefont {Xie}},
\bibinfo {author}
	{\bibfnamefont {R.}~\bibnamefont {Bidthanapally}},
\bibinfo {author}
	{\bibfnamefont {J.}~\bibnamefont {Sklenar}},
\bibinfo {author}
	{\bibfnamefont {P.}~\bibnamefont {Li}},
\bibinfo {author}
	{\bibfnamefont {S.}~\bibnamefont {Louis}},
\bibinfo {author}
	{\bibfnamefont {V.}~\bibnamefont {Tyberkevych}},
\bibinfo {author}
	{\bibfnamefont {H.}~\bibnamefont {Qu}}, \emph {et~al.},\ }
\bibfield  {title}
{\bibinfo {title}
	{Tunable magnetically induced transparency spectra in magnon-magnon coupled Y$_3$Fe$_5$O$_{12}$/permalloy bilayers},\ }\href@noop {}
{\bibfield  {journal}
	{\bibinfo  {journal} {Phys. Rev. Appl.}\ }\textbf {\bibinfo
		{volume} {17}},\ \bibinfo {pages} {044010} (\bibinfo {year} {2022})}\BibitemShut
{NoStop}%
%ref32
\bibitem [{\citenamefont {Wang}\ \emph {et~al.}(2019)\citenamefont {Wang},
	\citenamefont {Rao}, \citenamefont {Yang}, \citenamefont {Xu},\citenamefont {Gui},\citenamefont {Yao},\citenamefont {You},\ and\ \citenamefont {Hu}}]{wang2019nonreciprocity}%
\BibitemOpen
\bibfield  {author}
{\bibinfo {author}
	{\bibfnamefont {Y.-P.}~\bibnamefont {Wang}},
\bibinfo {author}
	{\bibfnamefont {J.}~\bibnamefont {Rao}},
\bibinfo {author}
	{\bibfnamefont {Y.}~\bibnamefont {Yang}},
\bibinfo {author}
	{\bibfnamefont {P.-C.}~\bibnamefont {Xu}},
\bibinfo {author}
	{\bibfnamefont {Y.}~\bibnamefont {Gui}},
\bibinfo {author}
	{\bibfnamefont {B.}~\bibnamefont {Yao}},
\bibinfo {author}
	{\bibfnamefont {J.Q.}~\bibnamefont {You}},
	\ and\
\bibinfo {author}
	{\bibfnamefont {C.-M.}~\bibnamefont {Hu}},\ }
\bibfield  {title}
{\bibinfo {title}
	{Nonreciprocity and unidirectional invisibility in cavity magnonics},\ }\href@noop {}
{\bibfield  {journal}
	{\bibinfo  {journal} {Phys. Rev. Lett.}\ }\textbf {\bibinfo
		{volume} {123}},\ \bibinfo {pages} {127202} (\bibinfo {year} {2019})}\BibitemShut
{NoStop}%
%ref33
\bibitem [{\citenamefont {Zhang}\ \emph {et~al.}(2020)\citenamefont {Zhang},\citenamefont {Galda},\citenamefont {Han},\citenamefont {Jin},\citenamefont {Nori},\citenamefont {Liu}, \ and\ \citenamefont {Vinokur}}]{zhang2020broadband}%
\BibitemOpen
\bibfield  {author}
{\bibinfo {author}
	{\bibfnamefont {X.}~\bibnamefont {Zhang}},
\bibinfo {author}
	{\bibfnamefont {A.}~\bibnamefont {Galda}},
\bibinfo {author}
	{\bibfnamefont {X.}~\bibnamefont {Han}},
\bibinfo {author}
	{\bibfnamefont {D.}~\bibnamefont {Jin}},
	\ and\
\bibinfo {author}
	{\bibfnamefont {V.}~\bibnamefont {Vinokur}},\ }
\bibfield  {title}
{\bibinfo {title}
	{Broadband nonreciprocity enabled by strong coupling of magnons and microwave photons},\ }\href@noop {}
{\bibfield  {journal}
	{\bibinfo  {journal} {Phys. Rev. Appl.}\ }\textbf {\bibinfo
		{volume} {13}},\ \bibinfo {pages} {044039} (\bibinfo {year} {2020})}\BibitemShut
{NoStop}%
%ref34
\bibitem [{\citenamefont {Li}\ \emph {et~al.}(2023)\citenamefont {Li},
	\citenamefont {Lo}, \citenamefont {Lim}, \citenamefont {Pearson},\citenamefont {Divan},\citenamefont {Zhang},\citenamefont {Welp},\citenamefont {Kwok},\citenamefont {Hoﬀmann},\ and\ \citenamefont {Novosad}}]{li2023unidirectional}%
\BibitemOpen
\bibfield  {author}
{\bibinfo {author}
	{\bibfnamefont {Y.}~\bibnamefont {Li}},
\bibinfo {author}
	{\bibfnamefont {T.-H. }~\bibnamefont {Lo}},
\bibinfo {author}
	{\bibfnamefont {J.}~\bibnamefont {Lim}},
\bibinfo {author}
	{\bibfnamefont {J. E.}~\bibnamefont {Pearson}},
\bibinfo {author}
	{\bibfnamefont {R.}~\bibnamefont {Divan}},
\bibinfo {author}
	{\bibfnamefont {W.}~\bibnamefont {Zhang}},
\bibinfo {author}
	{\bibfnamefont {U.}~\bibnamefont {Welp}},
\bibinfo {author}
	{\bibfnamefont {W.-K.}~\bibnamefont {Kwok}},
\bibinfo {author}
	{\bibfnamefont {A.}~\bibnamefont {Hoffmann}},
	\ and\
\bibinfo {author}
	{\bibfnamefont {V.}~\bibnamefont {Novosad}},\ }
\bibfield  {title}
{\bibinfo {title}
	{Unidirectional microwave transduction with chirality selected short-wavelength magnon excitations},\ }\href@noop {}
{\bibfield  {journal}
	{\bibinfo  {journal} {Appl. Phys. Lett}\ }\textbf {\bibinfo
		{volume} {123}},\ \bibinfo {pages} {2} (\bibinfo {year} {2023})}\BibitemShut
{NoStop}%	
 %ref35
 \bibitem [{\citenamefont {Harder}\ \emph {et~al.}(2018)\citenamefont {Harder},
 	\citenamefont {Yang}, \citenamefont {Yao}, \citenamefont {Yu},\citenamefont {Rao},\citenamefont {Gui},\citenamefont {Stamps},\ and\ \citenamefont {Hu}}]{harder2018level}%
 \BibitemOpen
 \bibfield  {author}
 {\bibinfo {author}
 	{\bibfnamefont {M.}~\bibnamefont {Harder}},
 \bibinfo {author}
 	{\bibfnamefont {Y.}~\bibnamefont {Yang}},
 \bibinfo {author}
 	{\bibfnamefont {B.}~\bibnamefont {Yao}},
 \bibinfo {author}
 	{\bibfnamefont {C.}~\bibnamefont {Yu}},
 \bibinfo {author}
 	{\bibfnamefont {J.}~\bibnamefont {Rao}},
 \bibinfo {author}
 	{\bibfnamefont {Y.}~\bibnamefont {Gui}},
 \bibinfo {author}
 	{\bibfnamefont {R.}~\bibnamefont {Stamps}},
 	\ and\
 \bibinfo {author}
 	{\bibfnamefont {C.-M.}~\bibnamefont {Hu}},\ }
 \bibfield  {title}
 {\bibinfo {title}
 	{Level attraction due to dissipative magnon-photon coupling},\ }\href@noop {}
 {\bibfield  {journal}
 	{\bibinfo  {journal} {Phys. Rev. Lett.}\ }\textbf {\bibinfo
 		{volume} {121}},\ \bibinfo {pages} {137203} (\bibinfo {year} {2018})}\BibitemShut
 {NoStop}%
 %ref36
 \bibitem [{\citenamefont {Boventer}\ \emph {et~al.}(2019)\citenamefont {Boventer}, \citenamefont {Kläui}, \citenamefont {Macêdo},\ and\ \citenamefont {Weides}}]{boventer2019steering}%
 \BibitemOpen
 \bibfield  {author}
 {\bibinfo {author}
 	{\bibfnamefont {I.}~\bibnamefont {Boventer}},
 	\bibinfo {author}
 	{\bibfnamefont {M.}~\bibnamefont {Kl{\"a}ui}},
 	\bibinfo {author}
 	{\bibfnamefont {R.}~\bibnamefont {Mac{\^e}do}},
 	\ and\ \bibinfo {author}
 	{\bibfnamefont {M.}~\bibnamefont {Weides}},\ }
 \bibfield  {title}
 {\bibinfo {title}
 	{Steering between level repulsion and attraction: broad tunability of two-port driven cavity magnon-polaritons},\ }\href@noop {}
 {\bibfield  {journal}
 	{\bibinfo  {journal} {New J. Phys.}\ }\textbf {\bibinfo
 		{volume} {21}},\ \bibinfo {pages} {125001} (\bibinfo {year} {2019})}\BibitemShut
 {NoStop}%
%ref37
\bibitem [{\citenamefont {Lachance}\ \emph {et~al.}(2019)\citenamefont {Lachance},\citenamefont {Tabuchi},\citenamefont {Gloppe},\citenamefont {Usami}, \ and\ \citenamefont {Nakamura}}]{lachance2019hybrid}%
\BibitemOpen
\bibfield  {author}
{\bibinfo {author}
	{\bibfnamefont {D.}~\bibnamefont {Lachance-Quirion}},
\bibinfo {author}
	{\bibfnamefont {Y.}~\bibnamefont {Tabuchi}},
\bibinfo {author}
	{\bibfnamefont {A.}~\bibnamefont {Gloppe}},
\bibinfo {author}
	{\bibfnamefont {K.}~\bibnamefont {Usami}},
	\ and\
\bibinfo {author}
	{\bibfnamefont {Y.}~\bibnamefont {Nakamura}},\ }
\bibfield  {title}
{\bibinfo {title}
	{Hybrid quantum systems based on magnonics},\ }\href@noop {}
{\bibfield  {journal}
	{\bibinfo  {journal} {Appl. Phys. Express}\ }\textbf {\bibinfo
		{volume} {12}},\ \bibinfo {pages} {070101} (\bibinfo {year} {2019})}\BibitemShut
{NoStop}%
%ref38
\bibitem [{\citenamefont {Wang}\ \emph {et~al.}(2020)\citenamefont {Wang}, \ and\ \citenamefont {Hu}}]{wang2020dissipative}%
\BibitemOpen
\bibfield  {author}
{\bibinfo {author}
	{\bibfnamefont {Y.-P.}~\bibnamefont {Wang}},
	\ and\
\bibinfo {author}
	{\bibfnamefont {C.-M.}~\bibnamefont {Hu}},\ }
\bibfield  {title}
{\bibinfo {title}
	{Dissipative couplings in cavity magnonics},\ }\href@noop {}
{\bibfield  {journal}
	{\bibinfo  {journal} {J. Appl. Phys.}\ }\textbf {\bibinfo
		{volume} {127}},\ \bibinfo {pages} {13} (\bibinfo {year} {2020})}\BibitemShut
{NoStop}%
%ref39
\bibitem [{\citenamefont {Li}\ \emph {et~al.}(2020)\citenamefont {Li},\citenamefont {Zhang},\citenamefont {Tyberkevych},\citenamefont {Kwok},\citenamefont {Hoffmann}, \ and\ \citenamefont {Novosad}}]{li2020hybrid}%
\BibitemOpen
\bibfield  {author}
{\bibinfo {author}
	{\bibfnamefont {Y.}~\bibnamefont {Li}},
\bibinfo {author}
	{\bibfnamefont {W.}~\bibnamefont {Zhang}},
\bibinfo {author}
	{\bibfnamefont {V.}~\bibnamefont {Tyberkevych}},
\bibinfo {author}
	{\bibfnamefont {W.-K.}~\bibnamefont {Kwok}},
\bibinfo {author}
	{\bibfnamefont {A.}~\bibnamefont {Hoffmann}},
	\ and\
\bibinfo {author}
	{\bibfnamefont {V.}~\bibnamefont {Novosad}},\ }
\bibfield  {title}
{\bibinfo {title}
	{Hybrid magnonics: physics, circuits, and applications for coherent information processing},\ }\href@noop {}
{\bibfield  {journal}
	{\bibinfo  {journal} {J. Appl. Phys.}\ }\textbf {\bibinfo
		{volume} {128}},\ \bibinfo {pages} {13} (\bibinfo {year} {2020})}\BibitemShut
{NoStop}%
%ref40
\bibitem [{\citenamefont {Harder}\ \emph {et~al.}(2021)\citenamefont {Harder},\citenamefont {Yao},\citenamefont {Gui}, \ and\ \citenamefont {Hu}}]{harder2021coherent}%
\BibitemOpen
\bibfield  {author}
{\bibinfo {author}
	{\bibfnamefont {M.}~\bibnamefont {Harder}},
\bibinfo {author}
	{\bibfnamefont {B.}~\bibnamefont {Yao}},
\bibinfo {author}
	{\bibfnamefont {Y.}~\bibnamefont {Gui}},
	\ and\
\bibinfo {author}
	{\bibfnamefont {C.-M.}~\bibnamefont {Hu}},\ }
\bibfield  {title}
{\bibinfo {title}
	{Coherent and dissipative cavity magnonics},\ }\href@noop {}
{\bibfield  {journal}
	{\bibinfo  {journal} {J. Appl. Phys.}\ }\textbf {\bibinfo
		{volume} {129}},\ \bibinfo {pages} {20} (\bibinfo {year} {2021})}\BibitemShut
{NoStop}%
%ref41
\bibitem [{\citenamefont {Yuan}\ \emph {et~al.}(2022)\citenamefont {Yuan},
	\citenamefont {Cao}, \citenamefont {Kamra}, \citenamefont {Duine},\ and\ \citenamefont {Yan}}]{yuan2022quantum}%
\BibitemOpen
\bibfield  {author}
{\bibinfo {author}
	{\bibfnamefont {H.}~\bibnamefont {Yuan}},
\bibinfo {author}
	{\bibfnamefont {Y. }~\bibnamefont {Cao}},
\bibinfo {author}
	{\bibfnamefont {A.}~\bibnamefont {Kamra}},
\bibinfo {author}
	{\bibfnamefont {R. A.}~\bibnamefont {Duine}},
	\ and\
\bibinfo {author}
	{\bibfnamefont {P.}~\bibnamefont {Yan}},\ }
\bibfield  {title}
{\bibinfo {title}
	{Quantum magnonics: when magnon spintronics meets quantum information science},\ }\href@noop {}
{\bibfield  {journal}
	{\bibinfo  {journal} {Phys. Rep.}\ }\textbf {\bibinfo
		{volume} {965}},\ \bibinfo {pages} {1} (\bibinfo {year} {2022})}\BibitemShut
{NoStop}%
%SM ref42
\bibitem [{Sup()}]{Sup}%
\BibitemOpen
\bibfield  {title} {\bibinfo {title} {See Supplemental Material [URL] for a detailed
		description of theoretical and experimental methods, which includes Refs. [43-50]}}\href@noop{}{}\BibitemShut {NoStop}%
%ref 43
	\bibitem [{\citenamefont {Wang}}(2018)]{wang2018time}%
	\BibitemOpen
	\bibfield  {author}
	{\bibinfo {author}
		{\bibfnamefont {K. X.}~\bibnamefont {Wang}},\ }
	\bibfield  {title}
	{\bibinfo {title}
		{Time-reversal symmetry in temporal coupled-mode theory and nonreciprocal device applications},\ }\href@noop {}
	{\bibfield  {journal}%
		{\bibinfo  {journal} {Opt. Lett.}\ }\textbf {\bibinfo {volume} {43}},\ \bibinfo {pages} {5623} (\bibinfo {year} {2018})}\BibitemShut
	{NoStop}%
%ref 44
	\bibitem [{\citenamefont {Zhao}\ \emph {et~al.}(2019)\citenamefont {Zhao},
		\citenamefont {Guo},\ and\ \citenamefont {Fan}}]{zhao2019connection}%
	\BibitemOpen
	\bibfield  {author}
	{\bibinfo {author}
		{\bibfnamefont {Z.}~\bibnamefont {Zhao}},
		\bibinfo {author}
		{\bibfnamefont {C.}~\bibnamefont {Guo}},\ and\
		\bibinfo {author}
		{\bibfnamefont {S.}~\bibnamefont {Fan}},\ }
	\bibfield  {title}
	{\bibinfo {title}
		{Connection of temporal coupled-mode-theory formalisms for a resonant optical system and its time-reversal conjugate},\ }\href@noop {}
	{\bibfield  {journal}%
		{\bibinfo  {journal} {Phys. Rev. A}\ }\textbf {\bibinfo {volume} {99}},\ \bibinfo {pages} {033839} (\bibinfo {year} {2019})}\BibitemShut
	{NoStop}%	
	%ref 45
\bibitem [{\citenamefont {Wang}\ and\ \citenamefont {Xiao}(2025)}]{wang2025interpreting}%
\BibitemOpen
\bibfield  {author}
{\bibinfo {author}
	{\bibfnamefont {J.}~\bibnamefont {Wang}}\ and\
	\bibinfo {author}
	{\bibfnamefont {J.}~\bibnamefont {Xiao}},\ }
\bibfield  {title}
{\bibinfo {title}
	{Interpreting S-parameter spectra in coupled resonant systems: the role of probing configurations},\ }\href@noop {}
{\bibfield  {journal}%
	{\bibinfo  {journal} {Opt. Express}\ }\textbf {\bibinfo {volume} {33}},\
	\bibinfo {pages} {7146} (\bibinfo {year} {2025})}\BibitemShut {NoStop}%
%ref 46
\bibitem [{\citenamefont {Yu}\ \emph {et~al.}(2020)\citenamefont {Yu},
	\citenamefont {Zhang}, \citenamefont {Sharma}, \citenamefont {Blanter},\ and\ \citenamefont {Bauer}}]{yu2020chiral}%
\BibitemOpen
\bibfield  {author}
{\bibinfo {author}
	{\bibfnamefont {T.}~\bibnamefont {Yu}},
	\bibinfo {author}
	{\bibfnamefont {X.}~\bibnamefont {Zhang}},
	\bibinfo {author}
	{\bibfnamefont {S.}~\bibnamefont {Sharma}},
	\bibinfo {author}
	{\bibfnamefont {Y. M.}~\bibnamefont {Blanter}},\ and\
	\bibinfo {author}
	{\bibfnamefont {G. E.}~\bibnamefont {Bauer}},\ }
\bibfield  {title}
{\bibinfo {title}
	{Chiral coupling of magnons in waveguides},\ }\href@noop {}
{\bibfield  {journal}%
	{\bibinfo  {journal} {Phys. Rev. B}\ }\textbf {\bibinfo {volume} {101}},\ \bibinfo {pages} {094414} (\bibinfo {year} {2020})}\BibitemShut
{NoStop}%
%ref 47
\bibitem [{\citenamefont {Yang}\ \emph {et~al.}(2024)\citenamefont {Yang},
	\citenamefont {Yao}, \citenamefont {Xiao}, \citenamefont {Fong},
	\citenamefont {Lau},\ and\ \citenamefont {Hu}}]{yang2024anomalous}%
\BibitemOpen
\bibfield  {author}
{\bibinfo {author}
	{\bibfnamefont {Y.}~\bibnamefont {Yang}},
	\bibinfo {author}
	{\bibfnamefont {J.}~\bibnamefont {Yao}},
	\bibinfo {author}
	{\bibfnamefont {Y.}~\bibnamefont {Xiao}},
	\bibinfo {author}
	{\bibfnamefont {P.-T.}~\bibnamefont {Fong}},
	\bibinfo {author}
	{\bibfnamefont {H.-K.}~\bibnamefont {Lau}},\ and\
	\bibinfo {author}
	{\bibfnamefont {C.-M.}~\bibnamefont {Hu}},\ }
\bibfield  {title}
{\bibinfo {title}
	{Anomalous long-distance coherence in critically driven cavity magnonics},\ }\href@noop {}
{\bibfield  {journal}%
	{\bibinfo  {journal} {Phys. Rev. Lett.}\ }\textbf {\bibinfo {volume} {132}},\ \bibinfo {pages} {206902} (\bibinfo {year} {2024})}\BibitemShut
{NoStop}
%ref 48
\bibitem [{\citenamefont {Haus}(1984)}]{haus1984waves}%
\BibitemOpen
\bibfield  {author}
{\bibinfo {author} {\bibfnamefont {H. A.}\ \bibnamefont
		{Haus}},\ }\href@noop {} {\emph {\bibinfo {title} {Waves and Fields in Optoelectronics}}}\
(\bibinfo  {publisher}
{Prentice-Hall, Englewood Cliﬀs, NJ},\ \bibinfo {year}
{1984})\BibitemShut
{NoStop}%
%ref 49
\bibitem [{\citenamefont {Teufel}\ \emph {et~al.}(2011)\citenamefont {Teufel},
	\citenamefont {Li}, \citenamefont {Allman}, \citenamefont {Cicak},
	\citenamefont {Sirois}, \citenamefont {Whittaker},\ and\ \citenamefont {Simmonds}}]{teufel2011circuit}%
\BibitemOpen
\bibfield  {author}
{\bibinfo {author}
	{\bibfnamefont {J. D.}~\bibnamefont {Teufel}},
	\bibinfo {author}
	{\bibfnamefont {D.}~\bibnamefont {Li}},
	\bibinfo {author}
	{\bibfnamefont {M. S.}~\bibnamefont {Allman}},
	\bibinfo {author}
	{\bibfnamefont {K.}~\bibnamefont {Cicak}},
	\bibinfo {author}
	{\bibfnamefont {A. J.}~\bibnamefont {Sirois}},
	\bibinfo {author}
	{\bibfnamefont {J. D.}~\bibnamefont {Whittaker}},\ and\
	\bibinfo {author}
	{\bibfnamefont {R. W.}~\bibnamefont {Simmonds}},\ }
\bibfield  {title}
{\bibinfo {title}
	{Circuit cavity electromechanics in the strong-coupling regime},\ }\href@noop {}
{\bibfield  {journal}%
	{\bibinfo  {journal} {Nature}\ }\textbf {\bibinfo {volume} {471}},\ \bibinfo {pages} {204} (\bibinfo {year} {2011})}\BibitemShut
{NoStop}%
%ref 50
\bibitem [{\citenamefont {Gehring}\ \emph {et~al.}(2003)\citenamefont {Gehring},
	\citenamefont {Schweinsberg}, \citenamefont {Barsi},\citenamefont {Kostinski}, \ and\ \citenamefont {Boyd}}]{gehring2006observation}%
\BibitemOpen
\bibfield  {author}
{\bibinfo {author}
	{\bibfnamefont {G. M.}~\bibnamefont {Gehring}},
	\bibinfo {author}
	{\bibfnamefont {A.}~\bibnamefont {Schweinsberg}},
	\bibinfo {author}
	{\bibfnamefont {C.}~\bibnamefont {Barsi}},
	\bibinfo {author}
	{\bibfnamefont {N.}~\bibnamefont {Kostinski}},
	\ and\
	\bibinfo {author}
	{\bibfnamefont {R. W.}~\bibnamefont {Boyd}},\ }
\bibfield  {title}
{\bibinfo {title}
	{Observation of backward pulse propagation through a medium with a negative group velocity},\ }\href@noop {}
{\bibfield  {journal}
	{\bibinfo  {journal} {Science}\ }\textbf {\bibinfo
		{volume} {312}},\ \bibinfo {pages} {895} (\bibinfo {year} {2006})}\BibitemShut
	 {NoStop}%
 %ref51
 \bibitem [{\citenamefont {Fan}\ \emph {et~al.}(2003)\citenamefont {Fan},\citenamefont {Suh}, \ and\ \citenamefont {Joannopoulos}}]{fan2003temporal}%
 \BibitemOpen
 \bibfield  {author}
 {\bibinfo {author}
 	{\bibfnamefont {S.}~\bibnamefont {Fan}},
 \bibinfo {author}
 	{\bibfnamefont {W.}~\bibnamefont {Suh}},
 	\ and\
 \bibinfo {author}
 	{\bibfnamefont {J. D.}~\bibnamefont {Joannopoulos}},\ }
 \bibfield  {title}
 {\bibinfo {title}
 	{Temporal coupled-mode theory for the Fano resonance in optical resonators},\ }\href@noop {}
 {\bibfield  {journal}
 	{\bibinfo  {journal} {J. Opt. Soc. Am. A}\ }\textbf {\bibinfo
 		{volume} {20}},\ \bibinfo {pages} {569} (\bibinfo {year} {2003})}\BibitemShut
 {NoStop}%
  %ref52
 \bibitem [{\citenamefont {Autler }\ \emph {et~al.}(1955)\citenamefont {Autler}, \ and\ \citenamefont {Townes}}]{autler1955stark}%
 \BibitemOpen
 \bibfield  {author}
 {\bibinfo {author}
 	{\bibfnamefont {S. H.}~\bibnamefont {Autler}},
 	\ and\
 \bibinfo {author}
 	{\bibfnamefont {C. H.}~\bibnamefont {Townes}},\ }
 \bibfield  {title}
 {\bibinfo {title}
 	{Stark effect in rapidly varying fields},\ }\href@noop {}
 {\bibfield  {journal}
 	{\bibinfo  {journal} {Phys. Rev.}\ }\textbf {\bibinfo
 		{volume} {100}},\ \bibinfo {pages} {703} (\bibinfo {year} {1955})}\BibitemShut
 {NoStop}%
 %ref53
 \bibitem [{\citenamefont {Peng}\ \emph {et~al.}(2014)\citenamefont {Peng},\citenamefont {\"{O}zdemir},\citenamefont {Chen},\citenamefont {Nori},
 	\ and\ \citenamefont {Yang}}]{peng2014and}%
 \BibitemOpen
 \bibfield  {author}
 {\bibinfo {author}
 	{\bibfnamefont {B.}~\bibnamefont {Peng}},
 \bibinfo {author}
 	{\bibfnamefont {\c{S}. K.}~\bibnamefont {\"{O}zdemir}},
 \bibinfo {author}
 	{\bibfnamefont {W.}~\bibnamefont {Chen}},
 \bibinfo {author}
 	{\bibfnamefont {F.}~\bibnamefont {Nori}}, \ and\
 \bibinfo {author}
 	{\bibfnamefont {L.}~\bibnamefont {Yang}},\ }
 \bibfield {title}
 {\bibinfo {title} {What is and what is not electromagnetically induced transparency in whispering-gallery microcavities},\ }\href@noop {}
 {\bibfield  {journal}
 	{\bibinfo  {journal}
 		{Nat. Commun.}\ }\textbf
 	{\bibinfo {volume}{5}},\ \bibinfo {pages} {5082} (\bibinfo {year} {2014})}\BibitemShut
 {NoStop}%
 %ref54
 \bibitem [{\citenamefont {Artman}\ \emph {et~al.}(1955)\citenamefont {Artman}, \ and\ \citenamefont {Tannenwald}}]{artman1955measurement}%
 \BibitemOpen
 \bibfield  {author}
 {\bibinfo {author}
 	{\bibfnamefont {J.}~\bibnamefont {Artman}},
 	\ and\
 \bibinfo {author}
 	{\bibfnamefont {P.}~\bibnamefont {Tannenwald}},\ }
 \bibfield  {title}
 {\bibinfo {title}
 	{Measurement of susceptibility tensor in ferrites},\ }\href@noop {}
 {\bibfield  {journal}
 	{\bibinfo  {journal} {J. Appl. Phys.}\ }\textbf {\bibinfo
 		{volume} {26}},\ \bibinfo {pages} {1124} (\bibinfo {year} {1955})}\BibitemShut
 {NoStop}%
 %ref55
 \bibitem [{\citenamefont {Huebl}\ \emph {et~al.}(2013)\citenamefont {Huebl},
 	\citenamefont {Zollitsch}, \citenamefont {Lotze}, \citenamefont {Hocke},\citenamefont {Greifenstein},\citenamefont {Marx},\citenamefont {Gross},\ and\ \citenamefont {Goennenwein}}]{huebl2013high}%
 \BibitemOpen
 \bibfield  {author}
 {\bibinfo {author}
 	{\bibfnamefont {H.}~\bibnamefont {Huebl}},
 \bibinfo {author}
 	{\bibfnamefont {C. W.}~\bibnamefont {Zollitsch}},
 \bibinfo {author}
 	{\bibfnamefont {J.}~\bibnamefont {Lotze}},
 \bibinfo {author}
 	{\bibfnamefont {F.}~\bibnamefont {Hocke}},
 \bibinfo {author}
 	{\bibfnamefont {M.}~\bibnamefont {Greifenstein}},
 \bibinfo {author}
 	{\bibfnamefont {A.}~\bibnamefont {Marx}},
 \bibinfo {author}
 	{\bibfnamefont {R.}~\bibnamefont {Gross}},
 	\ and\
 \bibinfo {author}
 	{\bibfnamefont {S. T.}~\bibnamefont {Goennenwein}},\ }
 \bibfield  {title}
 {\bibinfo {title}
 	{High cooperativity in coupled microwave resonator ferrimagnetic insulator hybrids},\ }\href@noop {}
 {\bibfield  {journal}
 	{\bibinfo  {journal} {Phys. Rev. Lett.}\ }\textbf {\bibinfo
 		{volume} {111}},\ \bibinfo {pages} {127003} (\bibinfo {year} {2013})}\BibitemShut
 {NoStop}%	
 %ref56
 \bibitem [{\citenamefont {Goryachev}\ \emph {et~al.}(2014)\citenamefont {Goryachev},\citenamefont {Farr},\citenamefont {Creedon},\citenamefont {Fan},\citenamefont {Kostylev}, \ and\ \citenamefont {Tobar}}]{goryachev2014high}%
 \BibitemOpen
 \bibfield  {author}
 {\bibinfo {author}
 	{\bibfnamefont {M.}~\bibnamefont {Goryachev}},
 \bibinfo {author}
 	{\bibfnamefont {W. G.}~\bibnamefont {Farr}},
 \bibinfo {author}
 	{\bibfnamefont {D. L.}~\bibnamefont {Creedon}},
 \bibinfo {author}
 	{\bibfnamefont {Y.}~\bibnamefont {Fan}},
 \bibinfo {author}
 	{\bibfnamefont {M.}~\bibnamefont {Kostylev}},
 	\ and\
 \bibinfo {author}
 	{\bibfnamefont {M. E.}~\bibnamefont {Tobar}},\ }
 \bibfield  {title}
 {\bibinfo {title}
 	{High-cooperativity cavity QED with magnons at microwave frequencies},\ }\href@noop {}
 {\bibfield  {journal}
 	{\bibinfo  {journal} {Phys. Rev. Appl.}\ }\textbf {\bibinfo
 		{volume} {2}},\ \bibinfo {pages} {054002} (\bibinfo {year} {2014})}\BibitemShut
 {NoStop}%
  %ref57
 \bibitem [{\citenamefont {Tabuchi}\ \emph {et~al.}(2014)\citenamefont {Tabuchi},
 	\citenamefont {Ishino}, \citenamefont {Ishikawa}, \citenamefont {Yamazaki},\citenamefont {Usami},\ and\ \citenamefont {Nakamura}}]{tabuchi2014hybridizing}%
 \BibitemOpen
 \bibfield  {author}
 {\bibinfo {author}
 	{\bibfnamefont {Y.}~\bibnamefont {Tabuchi}},
 \bibinfo {author}
 	{\bibfnamefont {S.}~\bibnamefont {Ishino}},
 \bibinfo {author}
 	{\bibfnamefont {T.}~\bibnamefont {Ishikawa}},
 \bibinfo {author}
 	{\bibfnamefont {R.}~\bibnamefont {Yamazaki}},
 \bibinfo {author}
 	{\bibfnamefont {K.}~\bibnamefont {Usami}},
 	\ and\
 \bibinfo {author}
 	{\bibfnamefont {Y.}~\bibnamefont {Nakamura}},\ }
 \bibfield  {title}
 {\bibinfo {title}
 	{Hybridizing ferromagnetic magnons and microwave photons in the quantum limit},\ }\href@noop {}
 {\bibfield  {journal}
 	{\bibinfo  {journal} {Phys. Rev. Lett.}\ }\textbf {\bibinfo
 		{volume} {113}},\ \bibinfo {pages} {083603} (\bibinfo {year} {2014})}\BibitemShut
 {NoStop}%	
   %ref58
 \bibitem [{\citenamefont {Bai}\ \emph {et~al.}(2015)\citenamefont {Bai},
 	\citenamefont {Harder}, \citenamefont {Chen}, \citenamefont {Fan},\citenamefont {Xiao},\ and\ \citenamefont {Hu}}]{bai2015spin}%
 \BibitemOpen
 \bibfield  {author}
 {\bibinfo {author}
 	{\bibfnamefont {L.}~\bibnamefont {Bai}},
 \bibinfo {author}
 	{\bibfnamefont {M.}~\bibnamefont {Harder}},
 \bibinfo {author}
 	{\bibfnamefont {Y.}~\bibnamefont {Chen}},
 \bibinfo {author}
 	{\bibfnamefont {X.}~\bibnamefont {Fan}},
 \bibinfo {author}
 	{\bibfnamefont {J.}~\bibnamefont {Xiao}},
 	\ and\
 \bibinfo {author}
 	{\bibfnamefont {C.-M.}~\bibnamefont {Hu}},\ }
 \bibfield  {title}
 {\bibinfo {title}
 	{Spin pumping in electrodynamically coupled magnon-photon systems},\ }\href@noop {}
 {\bibfield  {journal}
 	{\bibinfo  {journal} {Phys. Rev. Lett.}\ }\textbf {\bibinfo
 		{volume} {114}},\ \bibinfo {pages} {227201} (\bibinfo {year} {2015})}\BibitemShut
 {NoStop}%	
 %ref59
 \bibitem [{\citenamefont {Li}\ \emph {et~al.}(2019)\citenamefont {Li},\citenamefont    {Polakovic},\citenamefont {Wang},
 	\citenamefont {Xu}, \citenamefont {Lendinez}, \citenamefont {Zhang},\citenamefont {Ding},\citenamefont {Khaire},\citenamefont {Saglam},\citenamefont {Divan}\emph {et~al.}}]{li2019strong}%
 \BibitemOpen
 \bibfield  {author}
 {\bibinfo {author}
 	{\bibfnamefont {Y.}~\bibnamefont {Li}},
 \bibinfo {author}
 	{\bibfnamefont {T.}~\bibnamefont {Polakovic}},
 \bibinfo {author}
 	{\bibfnamefont {Y.-L. }~\bibnamefont {Wang}},
 \bibinfo {author}
 	{\bibfnamefont {J.}~\bibnamefont {Xu}},
 \bibinfo {author}
 	{\bibfnamefont {S.}~\bibnamefont {Lendinez}},
 \bibinfo {author}
 	{\bibfnamefont {Z.}~\bibnamefont {Zhang}},
 \bibinfo {author}
 	{\bibfnamefont {J.}~\bibnamefont {Ding}},
 \bibinfo {author}
 	{\bibfnamefont {T.}~\bibnamefont {Khaire}},
 \bibinfo {author}
 	{\bibfnamefont {H.}~\bibnamefont {Saglam}},
 \bibinfo {author}
 	{\bibfnamefont {R.}~\bibnamefont {Divan}}, \emph {et~al.},\ }
 \bibfield  {title}
 {\bibinfo {title}
 	{Strong coupling between magnons and microwave photons in on-chip ferromagnet-superconductor thin-film devices},\ }\href@noop {}
 {\bibfield  {journal}
 	{\bibinfo  {journal} {Phys. Rev. Lett.}\ }\textbf {\bibinfo
 		{volume} {123}},\ \bibinfo {pages} {107701} (\bibinfo {year} {2019})}\BibitemShut
 {NoStop}%
 %ref60
 \bibitem [{\citenamefont {Hou}\ \emph {et~al.}(2019)\citenamefont {Hou},
  \ and\ \citenamefont {Liu}}]{hou2019strong}%
 \BibitemOpen
 \bibfield  {author}
 {\bibinfo {author}
 	{\bibfnamefont {J. T.}~\bibnamefont {Hou}},
 	\ and\
 \bibinfo {author}
 	{\bibfnamefont {L.}~\bibnamefont {Liu}},\ }
 \bibfield  {title}
 {\bibinfo {title}
 	{Strong coupling between microwave photons and nanomagnet magnons},\ }\href@noop {}
 {\bibfield  {journal}
 	{\bibinfo  {journal} {Phys. Rev. Lett.}\ }\textbf {\bibinfo
 		{volume} {123}},\ \bibinfo {pages} {107702} (\bibinfo {year} {2019})}\BibitemShut
 {NoStop}%
  %ref61
 \bibitem [{Note1()}]{footnote}%
 \BibitemOpen
 \bibinfo {note} {Eq. \ref{Nonreciprocal EIT S21} is derived \cite{Sup} under the condition of $L \ll \lambda$, where $\lambda$ is the wavelength of the traveling wave. It predicts an additional relation of $\tau_{21}(\Delta_c)=\tau_{12}(-\Delta_c)$ as shown in Fig. 3(b), which is not preserved in our experiment performed at $L\approx 1.0$ cm and $\lambda \approx 3.3$ mm, see Fig. 3(a)}\BibitemShut
 {NoStop}%
 %ref62
 \bibitem [{\citenamefont {Kandic}\ \emph {et~al.}(2011)\citenamefont {Kandic},
 	\ and\ \citenamefont {Bridges}}]{kandic2011asymptotic}%
 \BibitemOpen
 \bibfield  {author}
 {\bibinfo {author}
 	{\bibfnamefont {M.}~\bibnamefont {Kandic}},
 	\ and\
 	\bibinfo {author}
 	{\bibfnamefont {G. E. }~\bibnamefont {Bridges}},\ }
 \bibfield  {title}
 {\bibinfo {title}
 	{Asymptotic limits of negative group delay in active resonator-based distributed circuits},\ }\href@noop {}
 {\bibfield  {journal}%
 	{\bibinfo  {journal} {IEEE Trans. Circuits Syst. I: Regul. Pap.}\ }\textbf {\bibinfo
 		{volume} {58}},\ \bibinfo {pages} {1727 } (\bibinfo {year} {2011})}\BibitemShut
 {NoStop}%
%ref63
\bibitem [{\citenamefont {Igarashi}\ \emph {et~al.}(2023)\citenamefont {Igarashi},
	\citenamefont {Zhang}, \citenamefont {Remy}, \citenamefont {D{\'\i}az},
	\citenamefont {Lin}, \citenamefont {Hohlfeld}, \citenamefont {Hehn},
	\citenamefont {Mangin}, \citenamefont {Gorchon},\ and\ \citenamefont {Malinowski}}]{igarashi2023optically}%
\BibitemOpen
\bibfield  {author}
{\bibinfo {author}
	{\bibfnamefont {J.}~\bibnamefont {Igarashi}},
	\bibinfo {author}
	{\bibfnamefont {W.}~\bibnamefont {Zhang}},
	\bibinfo {author}
	{\bibfnamefont {Q.}~\bibnamefont {Remy}},
	\bibinfo {author}
	{\bibfnamefont {E.}~\bibnamefont {D{\'\i}az}},
	\bibinfo {author}
	{\bibfnamefont {J.-X.}~\bibnamefont {Lin}},
	\bibinfo {author}
	{\bibfnamefont {J.}~\bibnamefont {Hohlfeld}},
	\bibinfo {author}
	{\bibfnamefont {M.}~\bibnamefont {Hehn}},
	\bibinfo {author}
	{\bibfnamefont {S.}~\bibnamefont {Mangin}},
	\bibinfo {author}
	{\bibfnamefont {J.}~\bibnamefont {Gorchon}},
	\ and\
	\bibinfo {author}
	{\bibfnamefont {G.}~\bibnamefont {Malinowski}},\ }
\bibfield  {title}
{\bibinfo {title}
	{Optically induced ultrafast magnetization switching in ferromagnetic spin valves},\ }\href@noop {}
{\bibfield  {journal}
	{\bibinfo  {journal} {Nat. Mater.}\ }\textbf {\bibinfo
		{volume} {22}},\ \bibinfo {pages} {725} (\bibinfo {year} {2023})}\BibitemShut
{NoStop}%
%ref64
\bibitem [{\citenamefont {Dainone}\ \emph {et~al.}(2024)\citenamefont {Dainone},
	\citenamefont {Prestes}, \citenamefont {Renucci}, \citenamefont {Bouch{\'e}},
	\citenamefont {Morassi}, \citenamefont {Devaux}, \citenamefont {Lindemann},
	\citenamefont {George}, \citenamefont {Jaffr{\`e}s}, \citenamefont {Lemaitre},
	\ and\ \citenamefont {others}}]{dainone2024controlling}%
\BibitemOpen
\bibfield  {author}
{\bibinfo {author}
	{\bibfnamefont {P. A.}~\bibnamefont {Dainone}},
	\bibinfo {author}
	{\bibfnamefont {N. F.}~\bibnamefont {Prestes}},
	\bibinfo {author}
	{\bibfnamefont {P.}~\bibnamefont {Renucci}},
	\bibinfo {author}
	{\bibfnamefont {A.}~\bibnamefont {Bouch{\'e}}},
	\bibinfo {author}
	{\bibfnamefont {M.}~\bibnamefont {Morassi}},
	\bibinfo {author}
	{\bibfnamefont {X.}~\bibnamefont {Devaux}},
	\bibinfo {author}
	{\bibfnamefont {M.}~\bibnamefont {Lindemann}},
	\bibinfo {author}
	{\bibfnamefont {J. M.}~\bibnamefont {George}},
	\bibinfo {author}
	{\bibfnamefont {H.}~\bibnamefont {Jaffr{\`e}s}},
	\bibinfo {author}
	{\bibfnamefont {A.}~\bibnamefont {Lemaitre}},
\emph {et~al.},\  }
\bibfield  {title}
{\bibinfo {title}
	{Controlling the helicity of light by electrical magnetization switching},\ }\href@noop {}
{\bibfield  {journal}
	{\bibinfo  {journal} {Nature}\ }\textbf {\bibinfo
		{volume} {627}},\ \bibinfo {pages} {783} (\bibinfo {year} {2024})}\BibitemShut
{NoStop}%
\end{thebibliography}
%apsrev4-2.bst 2019-01-14 (MD) hand-edited version of %apsrev4-1.bst

\end{document}